\documentclass[10pt,journal,epsfig,twoside]{IEEEtran}
\usepackage{graphicx}
\usepackage{subfigure,color,epstopdf}
\usepackage{amssymb}
\usepackage{cite,comment,setspace,varwidth}
\usepackage{amsmath}
\usepackage{bm,tcolorbox}
\usepackage{algorithm}
\usepackage{algpseudocode}
\makeatletter

\newcommand{\Rmnum}[1]{\expandafter\@slowromancap\romannumeral #1@}
\newcommand{\mv}[1]{\mbox{\boldmath{$ #1 $}}}
\newcommand\asteriskfill{\leavevmode\xleaders\hbox{$\ast$}\hfill\kern0pt}
\newcommand{\tabincell}[2]{\begin{tabular}{@{}#1@{}}#2\end{tabular}}
\makeatother
\newtheorem{fact}{Fact}

\newtheorem{definition}{Definition}

\begin{document}
\title{Multi-Beam Multi-Hop Routing for Intelligent Reflecting Surfaces Aided Massive MIMO}
\author{Weidong Mei and Rui Zhang, \IEEEmembership{Fellow, IEEE}
\thanks{Part of this work has been presented in IEEE International Conference on Communications, Montreal, Canada, 2021\cite{mei2020massive}.}
\thanks{The authors are with the Department of Electrical and Computer Engineering, National University of Singapore, Singapore 117583 (e-mails: \{wmei, elezhang\}@nus.edu.sg).}}
\markboth{IEEE TRANSACTIONS ON WIRELESS COMMUNICATIONS}{IEEE TRANSACTIONS ON WIRELESS COMMUNICATIONS}
\maketitle

\begin{abstract}
Intelligent reflecting surface (IRS) is envisioned to play a significant role in future wireless communication systems as an effective means of reconfiguring the radio signal propagation environment. In this paper, we study a new multi-IRS aided massive multiple-input multiple-output (MIMO) system, where a multi-antenna BS transmits independent messages to a set of remote single-antenna users using orthogonal beams that are subsequently reflected by different groups of IRSs via their respective multi-hop passive beamforming over pairwise line-of-sight (LoS) links. We aim to select optimal IRSs and their beam routing path for each of the users, along with the active/passive beamforming at the BS/IRSs, such that the minimum received signal power among all users is maximized. This problem is particularly difficult to solve due to a new type of path separation constraints for avoiding the IRS-reflected signal induced interference among different users. To tackle this difficulty, we first derive the optimal BS/IRS active/passive beamforming solutions based on their practical codebooks given the reflection paths. Then we show that the resultant multi-beam multi-hop routing problem can be recast as an equivalent graph-optimization problem, which is however NP-complete. To solve this challenging problem, we propose an efficient recursive algorithm to partially enumerate the feasible routing solutions, which is able to effectively balance the performance-complexity trade-off. Numerical results demonstrate that the proposed algorithm achieves near-optimal performance with low complexity and outperforms other benchmark schemes. Useful insights into the optimal multi-beam multi-hop routing design are also drawn under different setups of the multi-IRS aided massive MIMO network.
\end{abstract}
\begin{IEEEkeywords}
	Intelligent reflecting surface, massive MIMO, passive beamforming, multi-beam multi-hop routing, graph theory.
\end{IEEEkeywords}

\section{Introduction}
Wireless communication systems in the last decade have undergone a remarkable progress with various advanced technologies successfully implemented, such as adaptive modulation and coding, dynamic resource allocation, hybrid digital and analog beamforming, etc., which significantly enhanced their throughput and efficiency. However, existing wireless technologies were designed mainly to adapt to or compensate the random and time-varying wireless channels only, but have very limited control over them, thus leaving an ultimate barrier uncleared in achieving ultra-reliable and ultra-high-capacity wireless systems in the future. Recently, intelligent reflecting surface (IRS) has emerged as an appealing solution to tackle this issue. By dynamically tuning its large number of reflecting elements (or so-called passive beamforming), IRS is able to ``reconfigure'' wireless channels and refine their realizations and/or distributions\cite{wu2019towards,wu2020intelligent,basar2019wireless}, rather than adapting to them only in the traditional approach. In addition, IRS elements do not require transmit or receive radio frequency (RF) chains as they simply reflect the incident signal as a passive array, thus drastically reducing the hardware cost and energy consumption as compared to traditional active transceivers and relays. Thus, by efficiently integrating IRSs into future wireless networks, a quantum-leap improvement in capacity and energy efficiency is anticipated over today's wireless systems.

Due to the great potential of IRS, its performance has been recently studied in the literature under different wireless system setups, such as IRS-aided multi-antenna/multiple-input multiple-output (MIMO) system\cite{wu2019intelligent,zhang2019capacity}, massive MIMO system\cite{jamali2020intelligent,zhi2021statistical}, orthogonal frequency division multiplexing (OFDM) system\cite{yang2020intelligent,yang2020irs}, non-orthogonal multiple access (NOMA) system\cite{zheng2020intelligent,hou2020reconfigurable}, multi-cell network\cite{pan2020multicell,mei2020performance}, simultaneous wireless information and power transfer\cite{wu2020swipt,pan2020intelligent}, mobile edge computing\cite{jiang2019over,bai2020latency}, physical-layer security\cite{cui2019secure,yu2020robust}, unmanned aerial vehicle (UAV) communication\cite{lu2020enabling,fang2020joint}, and so on. However, all of these works consider one or multiple distributed IRSs, which assist in the wireless communication between the base station (BS) and users with only one single signal reflection by each IRS. This simplified approach, however, generally results in suboptimal performance. This is because by properly deploying IRSs, strong line-of-sight (LoS) channels can be achieved for inter-IRS links, which can provide more pronounced cooperative passive beamforming (CPB) gains over the conventional single-IRS assisted system. In addition, leveraging the multiple signal reflections by IRSs provides a higher path diversity to bypass the dense obstacles in a complex environment and thereby establish a blockage-free end-to-end link between two communication nodes, which generally has a stronger strength compared to other randomly scattered links between them that suffer multi-path fading.

Inspired by the above, the authors in \cite{han2020cooperative} first proposed a double-IRS system, where a single-antenna BS serves a single-antenna user through a double-reflection link with two cooperative IRSs deployed near the BS and user, respectively. It was shown in \cite{han2020cooperative} that this system provides a CPB gain that increases {\it quartically} with the total number of IRS reflecting elements, thus is significantly higher than the {\it quadratic} growth of the passive beamforming gain in the conventional single-IRS link. The authors in \cite{you2020wireless,zheng2021efficient,zheng2021double,dong2021double} further extended \cite{han2020cooperative} to address the more practical Rician fading channel and multi-antenna/multi-user setups. Specifically, the authors in \cite{you2020wireless} proposed two different channel estimation schemes for the double-IRS aided single-user system under arbitrary and LoS-dominant inter-IRS channels, respectively. In \cite{zheng2021efficient}, the authors studied the channel estimation problem in a more challenging double-IRS aided multi-user MIMO system with coexisting single- and double-reflection links. Furthermore, the passive beamforming optimization for the two IRSs under this system was studied in \cite{zheng2021double}. Finally, the authors in \cite{dong2021double} considered a secure double-IRS aided system and optimized the two IRSs' passive beamforming to maximize the secrecy rate. Despite of the above recent works, the general multi-IRS aided multi-user communication system with multi-hop (i.e., more than two hops) signal reflections has not been investigated in the literature yet. Under this general setup with more available IRSs in the network, different end-to-end LoS paths can be achieved between the BS and multiple remote users at the same time via multi-hop signal reflections by different groups of IRSs selected. This thus gives rise to a new cooperative {\it multi-beam multi-hop (MBMH) routing} design problem, where the selected IRSs and their beam-routing paths for different users are jointly optimized with the active/passive beamforming at the BS/IRSs to maximize the received signal power at all users. 
\begin{figure}[!t]
\centering
\includegraphics[width=3.4in]{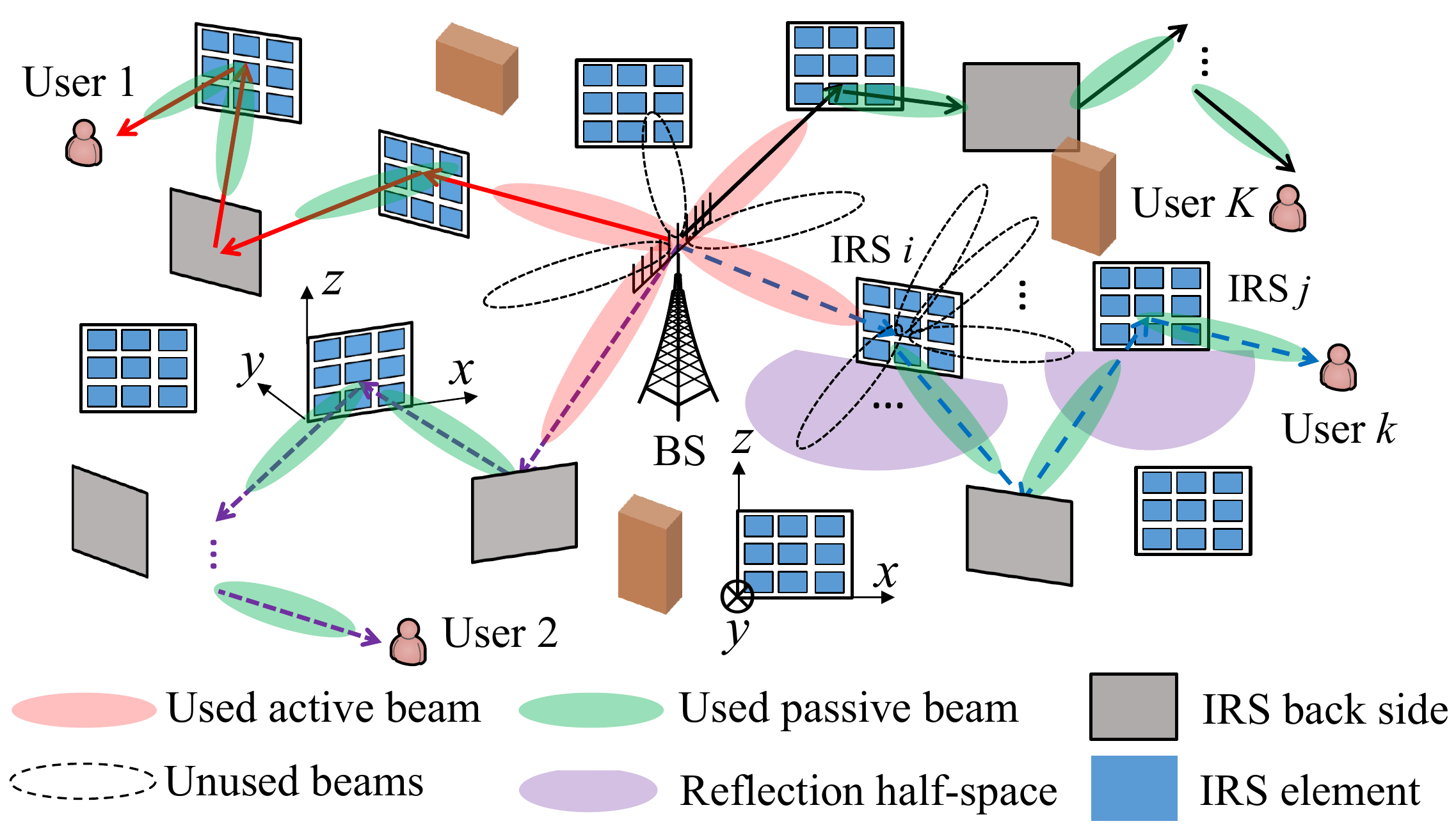}
\DeclareGraphicsExtensions.
\caption{A multi-IRS aided massive MIMO system with the MBMH routing via joint BS/IRS active/passive beamforming.}\label{MultiBeam}
\vspace{-12pt}
\end{figure}

\begin{table*}[htbp]
\centering
\small\caption{List of Main Symbols}\label{variable}
\begin{tabular}{|c|l|c|l|}
\hline
Symbol & Description & Symbol & Description\\
\hline
$J$ & Number of IRSs & $K$ & Number of users \\
\hline
$N_B$ & Number of BS antennas & $M$ & Number of IRS reflecting elements \\
\hline
${\cal W}_B$ & BS beamforming codebook, $\lvert{\cal W}_B\rvert=N_B$ & ${\cal K}$ & Set of users \\
\hline
${\cal J}$ & Set of IRSs & ${\mv \Phi}_j$ & Reflection coefficient matrix of IRS $j$ \\
\hline
${\mv \theta}_j$ & Passive beamforming vector of IRS $j$ & ${\cal W}_I$ & IRS beamforming codebook \\
\hline
$D$ & Number of beam patterns in ${\cal W}_I$, $D=\lvert{\cal W}_I\rvert$ & $b$ & Number of controlling bits for ${\cal W}_I$, $b=\log_2D$\\
\hline
${\mv H}_{0,j}$ & Channel from the BS to IRS $j$ & ${\mv S}_{i,j}$ & Channel from IRS $i$ to IRS $j$\\
\hline
${\mv g}^H_{j,J+k}$ & Channel from IRS $j$ to user $k$ & $d_A$/$d_I$ & Antenna/element spacing at BS/IRS\\
\hline
$M_1$/$M_2$ & \tabincell{l}{Number of IRS elements in horizontal/vertical\\ direction} & $d_{i,j}$ & Distance between nodes $i$ and $j$\\
\hline
$d_0$ & Minimum distance for far-field propagation & $l_{i,j}$ & LoS condition indicator between nodes $i$ and $j$\\
\hline
${\mv e}(\phi,N)$ & Steering vector function & $\lambda$ & Carrier wavelength\\
\hline
${\mv a}_B$ & Array response at the BS & ${\mv a}_I$ & Array response at each IRS\\
\hline
$\vartheta_{0,j}$ & AoD from the BS to IRS $j$ & $\varphi^a_{j,i}$/$\varphi^e_{j,i}$ & Azimuth/elevation AoA at IRS $j$ from node $i$\\
\hline
$\vartheta^a_{i,j}$/$\vartheta^e_{i,j}$ & Azimuth/elevation AoD from IRS $i$ to node $j$ & $\beta$ & LoS path gain at 1 meter\\
\hline
$\Omega^{(k)}$ & Reflection path from the BS to user $k$ & $N_k$ & Number of IRSs in $\Omega^{(k)}$\\
\hline
$a^{(k)}_n$ & Index of the $n$-th IRS in $\Omega^{(k)}$ & ${\cal N}_k$ & Set of IRSs in $\Omega^{(k)}$\\
\hline
$h_{0,J+k}(\Omega^{(k)})$ & BS-user $k$ effective channel & $\kappa(\Omega^{(k)})$ & End-to-end path gain between the BS and user $k$\\
\hline
${\mv w}_k$ & BS active beamforming for serving user $k$ & ${\cal W}_I^{(1)}$/${\cal W}_I^{(1)}$ & \tabincell{l}{IRS codebook for horizontal/vertical passive\\ beamforming}\\
\hline
$b_1$/$b_2$ & Number of controlling bits for ${\cal W}_I^{(1)}$/${\cal W}_I^{(1)}$ & $Q$ & Number of candidate shortest paths for each user\\
\hline
$G_p$ & Constructed path graph & $\Omega_r$ & Set of all cliques of size $r$ in $G_p$\\
\hline
\end{tabular}
\vspace{-6pt}
\end{table*}
In this paper, we study this new MBMH routing problem for the downlink communication in a massive MIMO system, where a BS equipped with a large number of active antennas transmits independent messages to a set of remote single-antenna users simultaneously over the same frequency band, aided by multiple distributed IRSs as shown in Fig.\,\ref{MultiBeam}. In \cite{mei2020cooperative}, by considering only a single user in this system, we have derived the optimal single-beam multi-hop routing solution. However, different from \cite{mei2020cooperative}, a new challenge arises in our considered MBMH routing design in this paper, which is to avoid the inter-user/path interference due to undesired scattering by the IRSs that serve for different users/paths, especially when there exist LoS channels between them. This thus leads to a new type of path separation constraints among different users, where the IRSs selected for different users/paths should avoid having LoS channels with each other. This stringent constraint thus makes the MBMH routing problem in this paper more challenging to solve, as compared to its single-beam special case in \cite{mei2020cooperative} without the inter-user/path interference considered. Moreover, unlike \cite{mei2020massive} and \cite{mei2020cooperative} where the continuous active/passive beamforming is assumed for the ease of exposition, in this paper we consider the more practical design based on beamforming codebook, which consists of only a finite number active/passive beamforming directions at the BS/IRSs, as shown in Fig.\,\ref{MultiBeam}. This helps reduce the complexity of optimal beamforming design as well as hardware cost in general, especially for frequency division duplex (FDD) systems.

To solve the proposed MBMH routing problem in a multi-IRS aided massive MIMO network, we first derive the optimal BS/IRS active/passive beamforming solution in their respective codebooks for given beam-routing paths of the users, by exploiting the high angular resolution of the massive MIMO BS and the inter-IRS LoS channels, respectively. Next, we show that the resultant MBMH routing problem is NP-complete by recasting it into an equivalent neighbor-disjoint path optimization problem in graph theory. To deal with this challenging problem, a recursive algorithm is proposed to partially enumerate the feasible MBMH routing solutions. By tuning its parameter, the proposed algorithm can strike a flexible balance between performance and complexity. It is also shown that in the special case of continuous passive beamforming at the IRSs, the MBMH routing problem can be solved in a more efficient manner by the proposed algorithm. Numerical results show that our proposed algorithm can find the near-optimal MBMH routing solution with low computational complexity and outperforms other benchmark schemes. It is also revealed that the optimal MBMH routing solution varies considerably with the number of reflecting elements as well as the size of passive beamforming codebook at each IRS. 

The rest of this paper is organized as follows. Section \Rmnum{2} presents the system model. Section \Rmnum{3} presents the optimal BS/IRS active/passive beamforming design and the problem formulation for our considered MBMH routing optimization. Section \Rmnum{4} presents the proposed solution to this problem based on graph theory. Section \Rmnum{5} presents the simulation results to show the performance of the proposed scheme as compared to other benchmark schemes. Finally, Section \Rmnum{7} concludes this paper and discusses future work.

The following notations are used in this paper. Bold symbols in capital letter and small letter denote matrices and vectors, respectively. The conjugate, transpose and conjugate transpose of a vector or matrix are denoted as ${(\cdot)}^{*}$, ${(\cdot)}^{T}$ and ${(\cdot)}^{H}$, respectively. ${\mathbb{R}}^n$ (${\mathbb{C}}^n$) denotes the set of real (complex) vectors of length $n$. For a complex number $s$, $s^*$ and $\lvert s \rvert$ denote its conjugate and amplitude, respectively. For a vector ${\mv a} \in {\mathbb{C}}^n$, ${\rm diag}({\mv a})$ denotes an $n \times n$ diagonal matrix whose entries are given by the elements of $\mv a$; while for a square matrix ${\mv A} \in {\mathbb{C}}^{n \times n}$, ${\rm diag}({\mv A})$ denotes an $n \times 1$ vector that contains the $n$ diagonal elements of ${\mv A}$. $\lVert \mv a \rVert$ denotes the Euclidean norm of the vector $\mv a$. $\lfloor \cdot \rfloor$ denotes the greatest integer less than or equal to its argument. $\lvert A \rvert$ denotes the cardinality of a set $A$. $j$ denotes the imaginary unit, i.e., $j^2=-1$. For two sets $A$ and $B$, $A \cup B$ denotes the union of $A$ and $B$. $\emptyset$ denotes an empty set. $\odot$ and $\otimes$ denote the Hadamard product and Kronecker product, respectively. ${\cal O}(\cdot)$ denotes the order of complexity. For ease of reference, the main symbols used in this paper are listed in Table \ref{variable}.

\section{System Model}
As shown in Fig.\,\ref{MultiBeam}, we consider a massive MIMO downlink system, where $J$ distributed IRSs are deployed to assist in the communications from a multi-antenna BS to $K$ remote single-antenna users. Assume that the BS is equipped with $N_B \gg K$ active antennas, while each IRS is equipped with $M$ passive reflecting elements. Without loss of generality and for ease of practical implementation, we assume that the BS serves the $K$ users by selecting $K$ beams from a predefined codebook, denoted as ${\cal W}_B$, which consists of $N_B$ orthogonal and unit-power beams, where $N_B$ can be arbitrarily large in massive MIMO. For the purpose of exposition, we consider the challenging scenario where the BS-user direct links are severely blocked for all the $K$ users considered in this paper. As such, the BS can only communicate with each user through a multi-reflection signal path that is formed by a set of IRSs associated with the user. To mitigate the potential inter-user interference during the multi-hop signal reflection, the signal paths for all $K$ users should be sufficiently separated and thus each IRS is associated with at most one user at one time, while it can serve multiple users over different time slots via proper user scheduling. As such, we focus on the MBMH routing design for a set of users in one given time slot. For convenience, we denote the sets of users and IRSs as ${\cal K}\triangleq \{1,2,\cdots,K\}$ and ${\cal J}\triangleq \{1,2,\cdots,J\}$, respectively. 

To maximize the reflected signal power by each selected IRS and ease the hardware implementation, we set the reflection amplitude of all its elements to the maximum value of one. As such, the reflection coefficient matrix of each IRS $j, j \in \cal J$ is given by ${\mv \Phi}_j={\rm diag}\{e^{j\theta_{j,1}},\cdots,e^{j\theta_{j,M}}\} \in {\mathbb C}^{M \times M}$, and its passive beamforming vector is denoted as ${\mv \theta}_j = {\rm diag}({\mv \Phi}_j) \in {\mathbb C}^{M \times 1}$. The passive beamforming vector of each IRS is assumed to be selected from a codebook ${\cal W}_I$, i.e., ${\mv \theta}_j \in {\cal W}_I, \forall j \in \cal J$, and ${\cal W}_I$ consists of $D=2^b$ beam patterns, where $b$ denotes the number of controlling bits for ${\cal W}_I$. For convenience, we refer to the BS and user $k, k \in \cal K$ as nodes 0 and $J+k$ in the system, respectively. Accordingly, we define ${\mv H}_{0,j} \in {\mathbb C}^{M \times N_B}, j \in {\cal J}$ as the channel from the BS to IRS $j$, ${\mv g}_{j,J+k}^{H} \in {\mathbb C}^{1 \times M}, j \in {\cal J}$ as that from IRS $j$ to user $k$, and ${\mv S}_{i,j} \in {\mathbb C}^{M \times M}, i,j \in {\cal J}, i \ne j$ as that from IRS $i$ to IRS $j$. For ease of exposition, we assume that the passive reflecting elements of each IRS in $\cal J$ are arranged in a uniform rectangular array (URA) perpendicular to the ground and facing a fixed direction, while the BS employs a uniform linear array (ULA). For convenience, we apply a three-dimensional (3D) coordinate system locally at each IRS and assume that its URA is parallel to the $x$-$z$ plane, as shown in Fig.\,\ref{MultiBeam}. The antenna and element spacing at the BS and each IRS is assumed to be $d_A$ and $d_I$, respectively. The numbers of elements in each IRS's horizontal and vertical directions are assumed to be $M_1$ and $M_2$, respectively, with $M_1M_2=M$. 

Let $d_{i,j}, i \ne j$ denote the distance between nodes $i$ and $j$, for which some reference transmitting/reflecting elements of the BS/IRSs are selected without loss of generality. To ensure the far-field propagation between any two nodes, we assume that $d_{i,j} \ge d_0, \forall i \ne j$, where $d_0$ denotes the minimum distance to satisfy this condition. According to \cite{han2020cooperative}, it must hold that $d_0 \gg \frac{{\sqrt M}d_I^2}{\lambda}$, where $\lambda$ denotes the carrier wavelength. Then, by carefully deploying the $J$ IRSs, LoS dominant propagation may be achieved between some pair of nodes $i$ and $j$ if $d_{i,j}$ is practically small (but larger than $d_0$). To simplify the active and passive beamforming designs as well as enhance the strength of the multi-reflection signal paths, we only exploit the LoS links in the system for the multi-hop signal reflection. Then, to describe the LoS condition between any two nodes $i$ (BS/IRS) and $j$ (IRS/user) in the considered system, we define a binary LoS condition indicator $l_{i,j} \in \{0,1\}$. In particular, $l_{i,j}=1$ indicates that the link between nodes $i$ and $j$ consists of an LoS link; otherwise, $l_{i,j}=0$. In addition, we set $l_{i,i}=0, \forall i$ and thus, $l_{i,j}=l_{j,i}, \forall i,j$. 

Furthermore, each IRS can only achieve 180$^\circ$ half-space reflection, i.e., only the signal incident on its reflection side can be reflected, as shown in Fig.\,\ref{MultiBeam}. Thus, for any two nodes $i$ and $j$, if they are both IRSs, each of them needs to be located in the reflection half-space of the other to achieve effective signal reflection between them. For example, in Fig.\,\ref{MultiBeam}, IRS $j$ and IRS $i$ cannot successively reflect the signal from the BS as they do not meet the above condition. Similarly, if one of the two nodes (say, node $i$) is the BS/user and the other node (say, node $j$) is an IRS, then node $i$ should be located in the reflection half-space of node $j$. Equivalently, if the above conditions cannot be satisfied for any two nodes $i$ and $j$, we can set $l_{i,j}=0$. In this paper, to focus on the new MBMH routing design, we assume that the LoS condition indicators $l_{i,j}$'s are known and constant after deploying the IRSs, while how to acquire such knowledge in practice is an interesting problem to be addressed in our future work. Based on the LoS condition between any two nodes in the considered system, a multi-hop LoS link can be established between the BS and each user $k, k \in \cal K$ by properly selecting a subset of associated IRSs. For example, if $l_{0,i}=l_{i,j}=l_{j,J+k}=1, i,j \in \cal J$, we can select IRSs $i$ and $j$ as the associated IRSs of user $k$, which successively reflect its intended signal from the BS toward its receiver. For all IRSs that are not associated with any user in $\cal K$, the BS can inform their controllers can via the control links to turn them off based on its optimized MBMH routing solution, so as to minimize the scattered interference in the system.

Next, we characterize the LoS channel between any two nodes in the system (if any), which is modeled as the product of array responses at their two sides. For convenience, we define the following steering vector function,
\begin{equation}\label{steer}
	{\mv e}(\phi,N)=[1,e^{-j\pi\phi},\cdots,e^{-j\pi(N-1)\phi}]^T \in {\mathbb C}^{N \times 1}, 
\end{equation}
where $N$ denotes the number of elements in a ULA, and $\phi$ denotes the phase difference between the observations at two adjacent elements. Obviously, ${\mv e}(\phi,N)$ is a periodic functions of $\phi$ and has a period of 2. Hence, we restrict $\phi \in [0,2)$ in the sequel of this paper. If $\phi \ge 2$ or $\phi < 0$, we set $\phi$ as $\phi - 2\lfloor \frac{\phi}{2} \rfloor$. Then, the array response at the BS is expressed as 
\begin{equation}\label{array1}
	{\mv a}_B(\vartheta)={\mv e}\Big(\frac{2d_A}{\lambda}\sin \vartheta,N_B\Big), 
\end{equation}
where $\vartheta$ denotes the angle-of-departure (AoD) relative to the BS antenna boresight. For the URA at each IRS, its array response is expressed as the Kronecker product of two steering vector functions in the horizontal and vertical directions, respectively, i.e.,
\begin{equation}\label{array2}
	{\mv a}_I(\vartheta^a,\vartheta^e)\!=\!{\mv e}\Big(\frac{2d_I}{\lambda}\sin\vartheta^e\cos\vartheta^a,M_1\Big) \otimes {\mv e}\Big(\frac{2d_I}{\lambda}\cos\vartheta^e,M_2\Big), 
\end{equation}
where $\vartheta^e$ and $\vartheta^a$ denote its elevation angle-of-arrival (AoA)/AoD and azimuth AoA/AoD, respectively. Then, we define $\vartheta_{0,j}$ as the AoD from the BS to IRS $j$, $\varphi^a_{j,i}$/$\varphi^e_{j,i}$ as the azimuth/elevation AoA at IRS $j$ from node $i$ (BS or IRS), and $\vartheta^a_{i,j}$/$\vartheta^e_{i,j}$ as the azimuth/elevation AoD from IRS $i$ to node $j$ (IRS or user). The above AoAs and AoDs can be estimated by exploiting the geometric relationship of the BS, IRSs and users in the system\cite{han2020cooperative} or by integrating sensors to the IRSs\cite{wu2020intelligent}.

Based on the above, we define ${\tilde{\mv h}}_{j,1}={\mv a}_B(\vartheta_{0,j})$ and ${\tilde{\mv h}}_{j,2}={\mv a}_I(\varphi^a_{j,0},\varphi^e_{j,0})$ for the LoS channel from the BS to IRS $j, j \in \cal J$, ${\tilde{\mv s}}_{i,j,1}={\mv a}_I(\vartheta^a_{i,j},\vartheta^e_{i,j})$ and ${\tilde{\mv s}}_{i,j,2}={\mv a}_I(\varphi^a_{j,i},\varphi^e_{j,i})$ for that from IRS $i$ to IRS $j, i,j \in \cal J$, as well as ${\tilde{\mv g}}_{j,J+k} = {\mv a}_I(\vartheta^a_{j,J+k},\vartheta^e_{j,J+k})$ for that from IRS $j$ to user $k, j \in {\cal J}, k \in \cal K$. Then, if $l_{0,j}=1$, the BS-IRS $j$ channel is expressed as
\begin{equation}\label{Ch1}
{\mv H}_{0,j} = \frac{\sqrt \beta}{d_{0,j}}e^{-\frac{j2\pi d_{0,j}}{\lambda}}{\tilde{\mv h}}_{j,2}{\tilde{\mv h}}^H_{j,1}, \;j \in {\cal J},
\end{equation}
where $\beta\, (<1)$ denotes the LoS path gain at the reference distance of 1 meter (m), and the exponential term captures the transmission delay over the LoS link. Similarly, if $l_{i,j}=1, i,j \in \cal J$, the IRS $i$-IRS $j$ channel is given by
\begin{equation}\label{Ch2}
{\mv S}_{i,j} = \frac{\sqrt \beta}{d_{i,j}}e^{-\frac{j2\pi d_{i,j}}{\lambda}}{\tilde{\mv s}}_{i,j,2}{\tilde{\mv s}}^H_{i,j,1}, \;i, j \in {\cal J}, i \ne j.
\end{equation}
Finally, if $l_{j,J+k}=1$, the IRS $j$-user $k$ channel is expressed as
\begin{equation}\label{Ch3}
{\mv g}^H_{j,J+k} \!=\! \frac{\sqrt \beta}{d_{j,J+k}}e^{-\frac{j2\pi d_{{j,J+k}}}{\lambda}}{\tilde{\mv g}}^H_{j,J+k}, \;j \!\in\! {\cal J}, k \!\in\! {\cal K}.
\end{equation}

Based on (\ref{Ch1})-(\ref{Ch3}), we can characterize the multi-hop LoS channel between the BS and each user $k, k \in \cal K$, with the given reflection path and BS/IRS active/passive beamforming. Specifically, let $\Omega^{(k)}=\{a^{(k)}_1,a^{(k)}_2,\cdots,a^{(k)}_{N_k}\}, k \in \cal K$ denote the reflection path from the BS to user $k$, where $N_k\, (\ge 1)$ and $a^{(k)}_n \in \cal J$ denote the number of associated IRSs for user $k$ and the index of the $n$-th associated IRS, with $n \in {\cal N}_k \triangleq \{1,2,\cdots,N_k\}$, respectively. For convenience, we define $a^{(k)}_0=0$ and $a^{(k)}_{N_k+1}=J+k, k \in \cal K$, corresponding to the BS and user $k$, respectively. Then, to ensure that each IRS in ${\cal N}_k$ only reflects user $k$'s information signal at most once, the following constraints should be met:
\begingroup
\allowdisplaybreaks
\begin{equation}\label{feasible1}
	a^{(k)}_n \in {\cal J}, \;a^{(k)}_n \ne a^{(k)}_{n'}, \forall n,n' \in {\cal N}_k, n \ne n', k \in {\cal K}. 
\end{equation}

Moreover, each constituent link of $\Omega^{(k)}$, along with the BS-IRS $a^{(k)}_1$ link and the IRS $a^{(k)}_{N_k}$-user $k$ link, should consist of an LoS link, i.e.,
\begin{equation}\label{feasible2}
	l_{a^{(k)}_n,a^{(k)}_{n+1}}=1, \forall n \in {\cal N}_k \cup \{0\}, k \in {\cal K}.
\end{equation}

Furthermore, to avoid the scattered inter-user interference, we consider that there is no direct LoS link\footnote{The methods and results in this paper are extendible to the more general path separation constraints, e.g., without $q$-hop LoS link between any two reflection paths, with $q \ge 1$, by utilizing a similar approach as in Section \ref{propsol}.} between any two nodes belonging to different reflection paths (except the common node $0$ or the BS). Thus, we have
\begin{equation}\label{feasible3}
	l_{a^{(k)}_n,a^{(k')}_{n'}}\!=\!0, \;a^{(k)}_n \!\ne\! a^{(k')}_{n'}, \forall n,n' \ne 0, k,k' \in {\cal K}, k \ne k'.
\end{equation}
Note that the condition $a^{(k)}_n \ne a^{(k')}_{n'}$ ensures that there is no common node (except the BS) between any two different reflection paths $\Omega^{(k)}$ and $\Omega^{(k')}$. As will be shown in Section \ref{sim}, the inter-user interference can be mitigated to a considerably lower level as compared to the information signal at each user $k$'s receiver thanks to the constraint (\ref{feasible3}).

Thus, each $\Omega^{(k)}$ is a feasible path if and only if the constraints in (\ref{feasible1})-(\ref{feasible3}) are satisfied. Given $K$ feasible paths $\Omega^{(k)}, k \in \cal K$, we define $\mv w_k \in {\mathbb C}^{N \times 1}, k \in \cal K$ as the BS active beamforming design for user $k$, with $\mv w_k \in {\cal W}_B$. Then, the BS-user $k$ effective channel is expressed as $h_{0,J+k}(\Omega^{(k)})=$
\begin{equation}\label{recvsig1}
{\mv g}^H_{a^{(k)}_{N_k},J+k}{\mv \Phi}_{a^{(k)}_{N_k}}\Big(\prod\limits_{n \in {\cal N}_k,\atop n \ne N_k}{\mv S}_{a^{(k)}_n,a^{(k)}_{n+1}}{\mv \Phi}_{a^{(k)}_n}\Big){\mv H}_{0,a^{(k)}_1}\mv w_k, \;k \in {\cal K},
\end{equation}
which depends on both the CPB design for the $N_k$ selected IRSs and the active beamforming design $\mv w_k$ for the BS. By substituting (\ref{Ch1})-(\ref{Ch3}) into (\ref{recvsig1}) and rearranging the terms in it, we obtain
\begin{equation}\label{recvsig2}
h_{0,J+k}(\Omega^{(k)})=e^{-j\varpi_k}\kappa(\Omega^{(k)})\Big(\prod\limits_{n=1}^{N_k}A^{(k)}_n\Big){\tilde{\mv h}}^H_{a^{(k)}_1,1}\mv w_k, k \in {\cal K},
\end{equation}
where
\begin{equation}\label{recvsig3}
A^{(k)}_n = \begin{cases}
	{\tilde{\mv s}}^H_{a^{(k)}_1,a^{(k)}_2,1}{\mv \Phi}_{a^{(k)}_1}{\tilde{\mv h}}_{a_1^{(k)},2} &{\text{if}}\;\;n=1\\
	{\tilde{\mv s}}^H_{a^{(k)}_n,a^{(k)}_{n+1},1}{\mv \Phi}_{a^{(k)}_n}{\tilde{\mv s}}_{a^{(k)}_{n-1},a^{(k)}_n,2}&{\text{if}}\;\;2 \le n \le N_k-1\\
	\tilde{\mv g}^H_{a^{(k)}_{N_k},J+k}{\mv \Phi}_{a^{(k)}_{N_k}}{\tilde{\mv s}}_{a^{(k)}_{N_k-1},a^{(k)}_{N_k},2} &{\text{if}}\;\;n=N_k,
\end{cases}
\end{equation}
$\varpi_k=\frac{2\pi}{\lambda}\sum\nolimits_{n=0}^{N_k}d_{a^{(k)}_n,a^{(k)}_{n+1}}$ is proportional to the end-to-end transmission distance, and
\begin{equation}\label{pathgain}
\kappa(\Omega^{(k)})=\frac{(\sqrt\beta)^{N_k+1}}{\prod\limits_{n=0}^{N_k}d_{a^{(k)}_n,a^{(k)}_{n+1}}}
\end{equation}
denotes the cascaded LoS path gain between the BS and user $k$ under the path $\Omega^{(k)}$, which turns out to be the product of the LoS path gains of all constituent links in $\Omega^{(k)}$. Note that the end-to-end delay under the path $\Omega^{(k)}$ is equal to $\sum\nolimits_{n=0}^{N_k}d_{a^{(k)}_n,a^{(k)}_{n+1}}$ divided by the speed of light, which is thus negligible.

Thus, the equivalent channel gain between the BS and user $k$, $\lvert h_{0,J+k}(\Omega^{(k)}) \rvert^2$, is expressed as
\begin{align}
\lvert h_{0,J+k}(\Omega^{(k)}) \rvert^2=\frac{\beta^{N_k+1}\prod\limits_{n=1}^{N_k} \lvert A^{(k)}_n \rvert^2 \cdot {\lvert\tilde{\mv h}}^H_{a_1^{(k)},1}\mv w_k\rvert^2}{\prod\limits_{n=0}^{N_k}d^2_{a^{(k)}_n,a^{(k)}_{n+1}}}, k \in {\cal K}.\label{eq1}
\end{align}

Based on (\ref{eq1}), we can obtain the optimal active/passive beamforming design at the BS/IRSs with a given MBMH routing solution, whereby the MBMH routing problem can be formulated, as detailed in the next section. 

\section{Optimal Beamforming Design and Problem Formulation}\label{bfpf}
In this section, we first derive the optimal active and passive beamforming design to maximize each $\lvert h_{0,J+k}(\Omega^{(k)}) \rvert^2, k \in \cal K$ in (\ref{eq1}) under a given reflection path $\Omega^{(k)}$. With the optimal beamforming design, we then formulate the MBMH routing problem where only the reflection paths $\Omega^{(k)}, k \in \cal K$, need to be optimized.\vspace{-6pt}

\subsection{Optimal Active and Passive Beamforming Design}\label{bf}
First, it is observed from (\ref{eq1}) that for any given reflection path for user $k$, to maximize $\lvert h_{0,J+k}(\Omega^{(k)}) \rvert^2$, the magnitude of each $A^{(k)}_n$ and ${\tilde{\mv h}}^H_{a_1^{(k)},1}\mv w_k$ should be maximized, subject to the codebook constraints at each IRS and the BS, respectively. First, given the codebook ${\cal W}_I$ at each IRS, consider that an IRS $j$ reflects the signal from its last node $i$ to the next node $r$. We denote by ${\mv \theta}_I(i,j,r)$ its corresponding optimal passive beamforming vector, which can be obtained by enumerating all beam patterns in ${\cal W}_I$, i.e., $\forall i,j,r$,\footnote{Similar passive beam search can also be performed in the case with mutual coupling among IRS elements, where the reflection coefficient matrix of each IRS is non-diagonal\cite{gradoni2021end}.}
\begin{equation}\label{theta1}
{\mv \theta}_I(i,j,r)=
\begin{cases}
	\arg \mathop{\max}\limits_{{\mv \theta} \in {\cal W}_I}\lvert{\tilde{\mv s}}^H_{j,r,1}{\rm diag}(\mv \theta){\tilde{\mv h}}_{j,2}\rvert\;&{\text{if}}\;i=0\\
	\arg \mathop{\max}\limits_{{\mv \theta} \in {\cal W}_I}\lvert\tilde{\mv g}^H_{j,J+k}{\rm diag}(\mv \theta){\tilde{\mv s}}_{i,j,2}\rvert \;&{\text{if}}\;r=J+k\\
	\arg \mathop{\max}\limits_{{\mv \theta} \in {\cal W}_I}\lvert{\tilde{\mv s}}^H_{j,r,1}{\rm diag}(\mv \theta){\tilde{\mv s}}_{i,j,2}\rvert\;&{\text{otherwise}}.
\end{cases}
\end{equation}

In particular, if the continuous passive beamforming with $b \rightarrow \infty$ is applied at each IRS, as all array responses have unit-modulus entries, (\ref{theta1}) can be simplified as
\begin{equation}\label{theta2}
{\mv \theta}_I(i,j,r)=
\begin{cases}
	\tilde{\mv s}_{j,r,1} \odot {\tilde{\mv h}}^*_{j,2} &{\text{if}}\;i=0\\
	\tilde{\mv g}_{j,J+k} \odot {\tilde{\mv s}}^*_{i,j,2} &{\text{if}}\;r=J+k\\
	\tilde{\mv s}_{j,r,1} \odot {\tilde{\mv s}}^*_{i,j,2}&{\text{otherwise}}.
	\end{cases}\;\forall i,j,r
\end{equation}

Accordingly, in the reflection path of user $k$, the passive beamforming of each IRS $a^{(k)}_n, n \in {\cal N}_k$ should be set as
\begin{equation}\label{search0}
{\mv \theta}_{a_n^{(k)}}={\rm diag}({\mv \Phi}_{a_n^{(k)}})={\mv \theta_I}(a^{(k)}_{n-1},a^{(k)}_n,a^{(k)}_{n+1}).
\end{equation}
Note that in the special case of continuous IRS beamforming with $b \rightarrow \infty$, by substituting (\ref{theta2}) and (\ref{search0}) into (\ref{recvsig3}), we have $A_n^{(k)}=M, \forall n \in {\cal N}_k, k \in \cal K$.\footnote{For ease of exposition, we assume that a full amplitude gain of $M$ can be obtained in this paper, while it may be dependent on the incident and reflection angles at each IRS in practice.}

To gain more useful insights into the optimal passive beamforming solutions at each IRS in (\ref{theta1}) and (\ref{theta2}), we consider that both nodes $i$ and $r$ are IRSs, which corresponds to the third case in (\ref{theta1}). Then, according to (\ref{array2}), it can be shown that\cite{you2020fast}
\begin{align}
&{\tilde{\mv s}}^H_{j,r,1}{\rm diag}(\mv \theta_j){\tilde{\mv s}}_{i,j,2}={\mv a}^H_I(\vartheta^a_{j,r},\vartheta^e_{j,r}){\rm diag}(\mv \theta_j){\mv a}_I(\varphi^a_{j,i},\varphi^e_{j,i})\nonumber\\
=&({\mv a}^H_I(\vartheta^a_{j,r},\vartheta^e_{j,r})\odot{\mv a}^T_I(\varphi^a_{j,i},\varphi^e_{j,i})){\mv \theta_j}\nonumber\\
=&\Big({\mv e}^H\Big(\frac{2d_I}{\lambda}\phi^{(1)}_{i,j,r},M_1\Big) \otimes {\mv e}^H\Big(\frac{2d_I}{\lambda}\phi^{(2)}_{i,j,r},M_2\Big)\Big){\mv \theta_j},\label{train1}
\end{align}
where $\phi^{(1)}_{i,j,r} \triangleq \sin\vartheta^e_{j,r}\cos\vartheta^a_{j,r}-\sin\varphi^e_{j,i}\cos\varphi^a_{j,i}$ and $\phi^{(2)}_{i,j,r} \triangleq \cos\vartheta^e_{j,r}-\cos\varphi^e_{j,i}$. Similarly, it can be verified that (\ref{train1}) also holds if node $i$ is the BS (the first case in (\ref{theta1})) or node $r$ is a user (the second case in (\ref{theta1})). It follows from (\ref{train1}) that if $b \rightarrow \infty$, the optimal passive beamforming in (\ref{theta2}) can be rewritten as ${\mv \theta}_I(i,j,r)={\mv e}\Big(\frac{2d_I}{\lambda}\phi^{(1)}_{i,j,r},M_1\Big) \otimes {\mv e}\Big(\frac{2d_I}{\lambda}\phi^{(2)}_{i,j,r},M_2\Big)$, which perfectly aligns the horizontal and vertical directions at the same time.

Motivated by this observation, we set ${\mv\theta}_j={\mv \theta}_I^{(1)}(i,j,r) \otimes {\mv \theta}_I^{(2)}(i,j,r)$ in (\ref{train1}), where ${\mv \theta}_I^{(1)}(i,j,r)$ and ${\mv \theta}_I^{(2)}(i,j,r)$ denote the horizontal and vertical passive beamforming vectors for IRS $j$, respectively. Then, we can obtain
\begin{align}
&{\tilde{\mv s}}^H_{j,r,1}{\rm diag}(\mv \theta_j){\tilde{\mv s}}_{i,j,2}\nonumber\\
=&\Big({\mv e}^H\Big(\frac{2d_I}{\lambda}\phi^{(1)}_{i,j,r},M_1\Big) \otimes {\mv e}^H\Big(\frac{2d_I}{\lambda}\phi^{(2)}_{i,j,r},M_2\Big)\Big)\!\cdot\!\Big({\mv \theta}_I^{(1)}(i,j,r)\nonumber\\
& \otimes {\mv \theta}_I^{(2)}(i,j,r)\Big)\nonumber\\
=&\Big({\mv e}^H\Big(\frac{2d_I}{\lambda}\phi^{(1)}_{i,j,r},M_1\Big){\mv \theta}_I^{(1)}(i,j,r)\Big)\cdot\Big({\mv e}^H\Big(\frac{2d_I}{\lambda}\phi^{(2)}_{i,j,r},M_2\Big)\nonumber\\
&{\mv \theta}_I^{(2)}(i,j,r)\Big).\label{train2}
\end{align}

It is noted from (\ref{train2}) that the passive beamforming of IRS $j$ can be decoupled into horizontal and vertical IRS passive beamforming. Accordingly, we define ${\cal W}^{(1)}_{I}$ and ${\cal W}^{(2)}_{I}$ as the codebooks for the horizontal and vertical IRS passive beamforming, respectively.\footnote{In the general case of multi-path channel model, a more sophisticated codebook structure may be needed. For example, in \cite{najafi2021physics}, an additional wavefront phase codebook is utilized to enable constructive or destructive superposition of the waves from different elements at the receivers.} The numbers of controlling bits for ${\cal W}^{(1)}_{I}$ and ${\cal W}^{(2)}_{I}$ are denoted as $b_1$ and $b_2$, respectively, which satisfy $b_1+b_2=b$. Hence, the IRS codebook ${\cal W}_{I}$ can be decomposed as
\begin{equation}\label{cb}
{\cal W}_{I} = \{{\mv \theta}|{\mv \theta}={\mv \theta}^{(1)} \otimes {\mv \theta}^{(2)}, {\mv \theta}^{(1)} \in {\cal W}^{(1)}_{I}, {\mv \theta}^{(2)} \in {\cal W}^{(2)}_{I}\},
\end{equation}
while the optimal IRS passive beamforming in (\ref{theta1}) can be computed as ${\mv \theta}_I(i,j,r)={\mv \theta}^{(1)}_I(i,j,r) \otimes {\mv \theta}^{(2)}_I(i,j,r)$, where
\begin{align}
{\mv \theta}^{(1)}_I(i,j,r) &= \arg \mathop{\max}\limits_{{\mv \theta}_1 \in {\cal W}^{(1)}_I}\left|{\mv e}^H\Big(\frac{2d_I}{\lambda}\phi^{(1)}_{i,j,r},M_1\Big){\mv \theta}_1\right|,\label{bs1}\\
{\mv \theta}^{(2)}_I(i,j,r) &= \arg \mathop{\max}\limits_{{\mv \theta}_2 \in {\cal W}^{(2)}_I}\left|{\mv e}^H\Big(\frac{2d_I}{\lambda}\phi^{(2)}_{i,j,r},M_2\Big){\mv \theta}_2\right|.\label{bs2}
\end{align}
Note that compared to the joint 3D beam search in (\ref{theta1}), the complexity of beam search can be greatly reduced from ${\cal O}(2^b)$ to ${\cal O}(2^{b_1}+2^{b_2})$ by separately solving (\ref{bs1}) and (\ref{bs2}). In particular, if node $r$ is not a user, i.e., $r \in \cal J$, the above beam search for any triple nodes $(i,j,r)$ can be conducted offline, since the BS and all IRSs are fixed and their channels can be assumed to be constant over a long period.

Accordingly, in the reflection path of user $k, k \in \cal K$, the passive beamforming of each IRS $a^{(k)}_n, n \in {\cal N}_k$ in (\ref{search0}) can be rewritten as
\begin{align}\label{search1}
{\mv \theta_I}(&a^{(k)}_{n-1},a^{(k)}_n,a^{(k)}_{n+1})=\nonumber\\
&{\mv \theta_I^{(1)}}(a^{(k)}_{n-1},a^{(k)}_n,a^{(k)}_{n+1}) \otimes {\mv \theta_I^{(2)}}(a^{(k)}_{n-1},a^{(k)}_n,a^{(k)}_{n+1}),
\end{align}
which can be simplified as
\begin{align}\label{search2}
	{\mv \theta_I}&(a^{(k)}_{n-1},a^{(k)}_n,a^{(k)}_{n+1})=\nonumber\\
	&{\mv e}\Big(\frac{2d_I}{\lambda}\phi^{(1)}_{a^{(k)}_{n-1},a^{(k)}_n,a^{(k)}_{n+1}},M_1\Big) \otimes
	{\mv e}\Big(\frac{2d_I}{\lambda}\phi^{(2)}_{a^{(k)}_{n-1},a^{(k)}_n,a^{(k)}_{n+1}},M_2\Big),
\end{align}
in the case of continuous passive beamforming at each IRS with $b_1, b_2 \rightarrow \infty$.

Next, we focus on the optimal active beamforming design for the BS, which should maximize the amplitude of ${\tilde{\mv h}}^H_{a_1^{(k)},1}\mv w_k, k \in \cal K$ in (\ref{eq1}). To this end, we define
\begin{equation}\label{search3}
{\mv w}_B(j)=\arg \mathop{\max}\limits_{{\mv w} \in {\cal W}_B} \lvert{\tilde{\mv h}}^H_{j,1}\mv w \rvert, j \in {\cal J},
\end{equation}
as the optimal active beamforming solution for the BS to transmit the beam to IRS $j$, which is obtained by enumerating all beam patterns in the codebook at the BS, ${\cal W}_B$, thus incurring the complexity of ${\cal O}(N_B)$. Similar to the beam search in (\ref{bs1}) and (\ref{bs2}), the beam search in (\ref{search3}) can also be performed offline. In particular, if the continuous active beamforming is applied at the BS, (\ref{search3}) becomes equivalent to the maximum-ratio transmission (MRT) based on ${\tilde{\mv h}}_{j,1}$, i.e.,
\begin{equation}\label{bmall}
	{\mv w}_B(j) = {\tilde{\mv h}}_{j,1}/{\lVert {\tilde{\mv h}}_{j,1} \rVert}, k \in {\cal K}.
\end{equation}
With this definition, the BS active beamforming in the reflection path of user $k$ should be set as
\begin{equation}\label{search4}
{\mv w}_k= {\mv w}_B(a^{(k)}_1)e^{j\varpi_k}, k \in {\cal K},
\end{equation}
where the effective phases $\varpi_k, k \in \cal K$ of the $K$ reflection paths are compensated at the BS.

It is worth noting that if $N_B$ is sufficiently large, the MRT-based beamforming in (\ref{bmall}) ensures that the power of the information signal for each user $k, k \in \cal K$ overwhelms that of the inter-user interference in the BS-IRS $\footnotesize {a^{(k)}_1}$ link, i.e., the first link in $\Omega^{(k)}$. This is because with a large $N_B$, the BS antenna array has a practically high angular resolution. If all first-hop IRSs in the reflection paths for the $K$ users, i.e., IRS $a^{(k)}_1, k \in \cal K$, are sufficiently separated in the angular domain, the following asymptotically favorable propagation for massive MIMO\cite{ngo2014aspects} can be achieved: 
\begin{equation}\label{eq2}
\begin{split}
&\frac{1}{N_B}{\lvert\tilde{\mv h}}^H_{a^{(k)}_1,1}{\mv w}_k\rvert^2 = 1, k \in {\cal K}, \\
&\frac{1}{N_B}{\lvert\tilde{\mv h}}^H_{a^{(k)}_1,1}{\mv w}_{k'}\rvert^2 \approx 0, k, k' \in {\cal K}, k \ne k'.
\end{split}
\end{equation}

The asymptotically favorable propagation in (\ref{eq2}) may also be achieved with practical finite-size codebooks, e.g., the discrete Fourier transform (DFT)-based codebook (see Section \ref{sim} for details). This is because when the codebook size $N_B$ is sufficiently large, the codebook will have a high resolution, such that the selected beam patterns are close to the MRT-based beamforming in (\ref{bmall}). Hence, the inter-user interference can be approximately nulled in the first link of each reflection path $\Omega^{(k)}$ by properly selecting ${\cal W}_B$. Furthermore, since the path separation constraints in (\ref{feasible3}) ensure that the scattered inter-user interference in the subsequent links of $\Omega^{(k)}$ is well mitigated, user $k$ is approximately free of inter-user interference, while achieving the maximum end-to-end channel gain with the BS via IRSs' passive beamforming in (\ref{search1}) and the BS's active beamforming in (\ref{search4}). 

Under the above optimal beamforming designs, we define $\tilde A^{(k)}_n$ as the maximum value of $A^{(k)}_n$ in (\ref{recvsig3}) by following (\ref{search1}). It is worth noting that $\tilde A^{(k)}_n$ depends on the AoAs/AoDs between nodes $a^{(k)}_{n-1}$ and $a^{(k)}_n$, as well as those between nodes $a^{(k)}_n$ and $a^{(k)}_{n+1}$. Besides, it also depends on the numbers of controlling bits for the IRS codebooks, i.e., $b_1$ and $b_2$. In particular, with increasing $b_1$ or $b_2$, the resolution of IRS codebook can be improved, thus resulting in a larger $\tilde A^{(k)}_n$. In the special case of continuous IRS beamforming with $b_1, b_2 \rightarrow \infty$, we have $\tilde A^{(k)}_n = M$, which is regardless of the AoAs/AoDs. It follows that the effect of AoAs and AoDs diminishes when the resolution of IRS codebooks becomes higher. By substituting (\ref{search1}) and (\ref{eq2}) into (\ref{eq1}), the maximum BS-user $k$ equivalent channel gain is given by
\begin{align}
\lvert h_{0,J+k}(\Omega^{(k)}) \rvert^2=\frac{\beta^{N_k+1}N_B\prod\limits_{n=1}^{N_k} \lvert \tilde A^{(k)}_n \rvert^2}{\prod\limits_{n=0}^{N_k}d^2_{a^{(k)}_n,a^{(k)}_{n+1}}}, k \in {\cal K}.\label{eq3}
\end{align}

It is observed from (\ref{eq3}) that besides the conventional active BS beamforming gain of $N_B$, a new multiplicative CPB gain of $\prod\nolimits_{n=1}^{N_k} \lvert \tilde A^{(k)}_n \rvert^2$ is also achieved for each BS-user $k$ equivalent channel. As previously discussed, if $b_1$ or $b_2$ is small, this multiplicative CPB gain will depend heavily on the AoAs and AoDs between any two consecutive nodes in $\Omega^{(k)}$. However, if both $b_1$ and $b_2$ are sufficiently large, this CPB gain can be greatly enhanced and approaches its maximum value, $M^{2N_k}$. In general, there exists a fundamental trade-off between maximizing the CPB gain versus the end-to-end path gain, i.e., $\kappa^2(\Omega^{(k)})$ in (\ref{pathgain}) (or minimizing the end-to-end path loss $\kappa^{-2}(\Omega^{(k)})$), as the former monotonically increases with $N_k$, while the latter generally decreases with $N_k$. Besides this trade-off, there exists another trade-off in balancing all $\lvert h_{0,J+k}(\Omega^{(k)}) \rvert$'s for different users in $\cal K$. Specifically, due to the practically finite number of IRSs and LoS paths in the system as well as the path separation constraints in (\ref{feasible3}), maximizing the channel gain for one user generally reduces the number of feasible paths for the other users. Particularly, if the number of users is large, some users may be denied access due to the lack of feasible paths. As such, the optimal MBMH routing design should reconcile the above trade-offs and take into account the resolution of practical IRS codebooks, so as to achieve the optimum performance of all $K$ users in a fair manner.

It should be mentioned that in addition to the reflection path $\Omega^{(k)}$, there may also exist some other signal paths between the BS and user $k$ due to the random scattering of all active IRSs in the system. Nonetheless, as will be shown in Section \ref{sim}, the strength of these scattered links is practically much lower than that of $\Omega^{(k)}$ in (\ref{eq3}), due to the lack of joint active and CPB gains over these scattered links. As such, we only focus on the multi-reflection path $\Omega_k$ in this paper.

\begin{figure}[!t]
\centering
\begin{minipage}[t]{0.49\textwidth}
\centering
\includegraphics[height=2.6in]{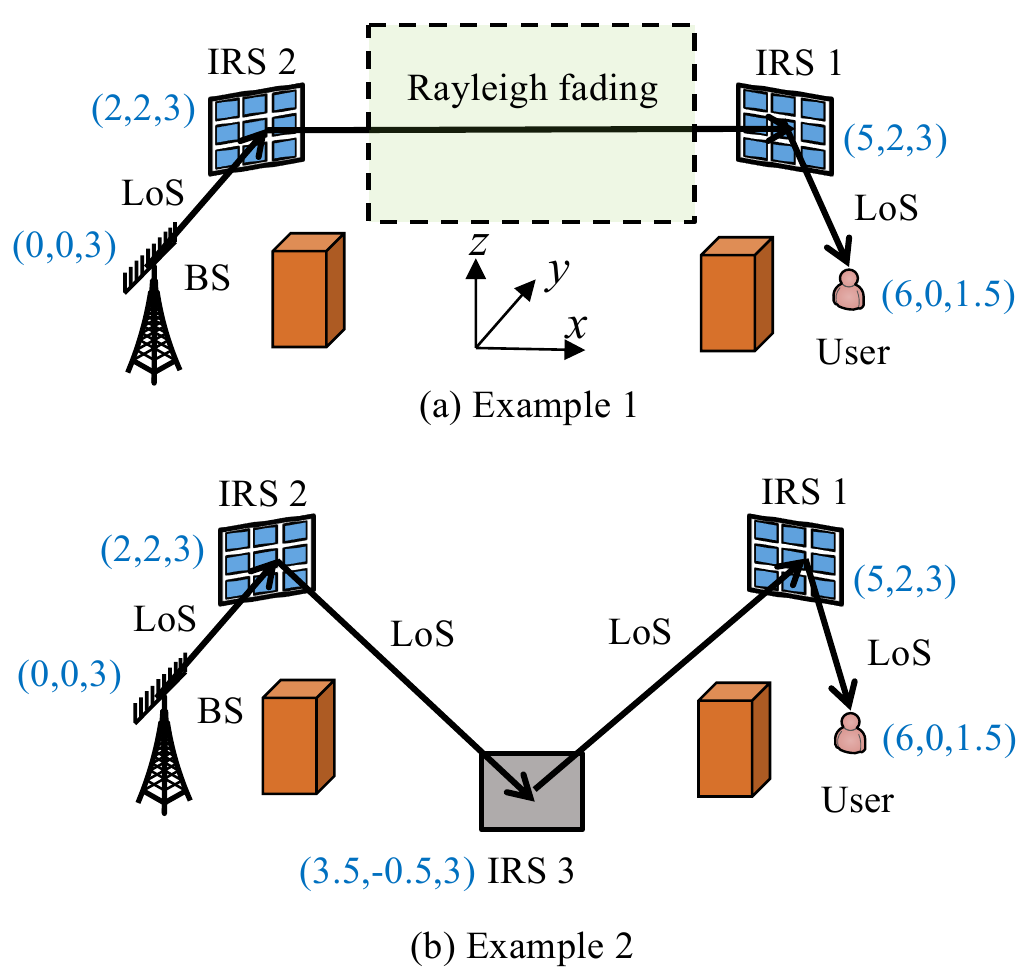}
\caption{Simulation setup of numerical examples.}\label{NumExpSetup}
\end{minipage}
\hfill
\begin{minipage}[t]{0.49\textwidth}
\centering
\includegraphics[height=2.6in]{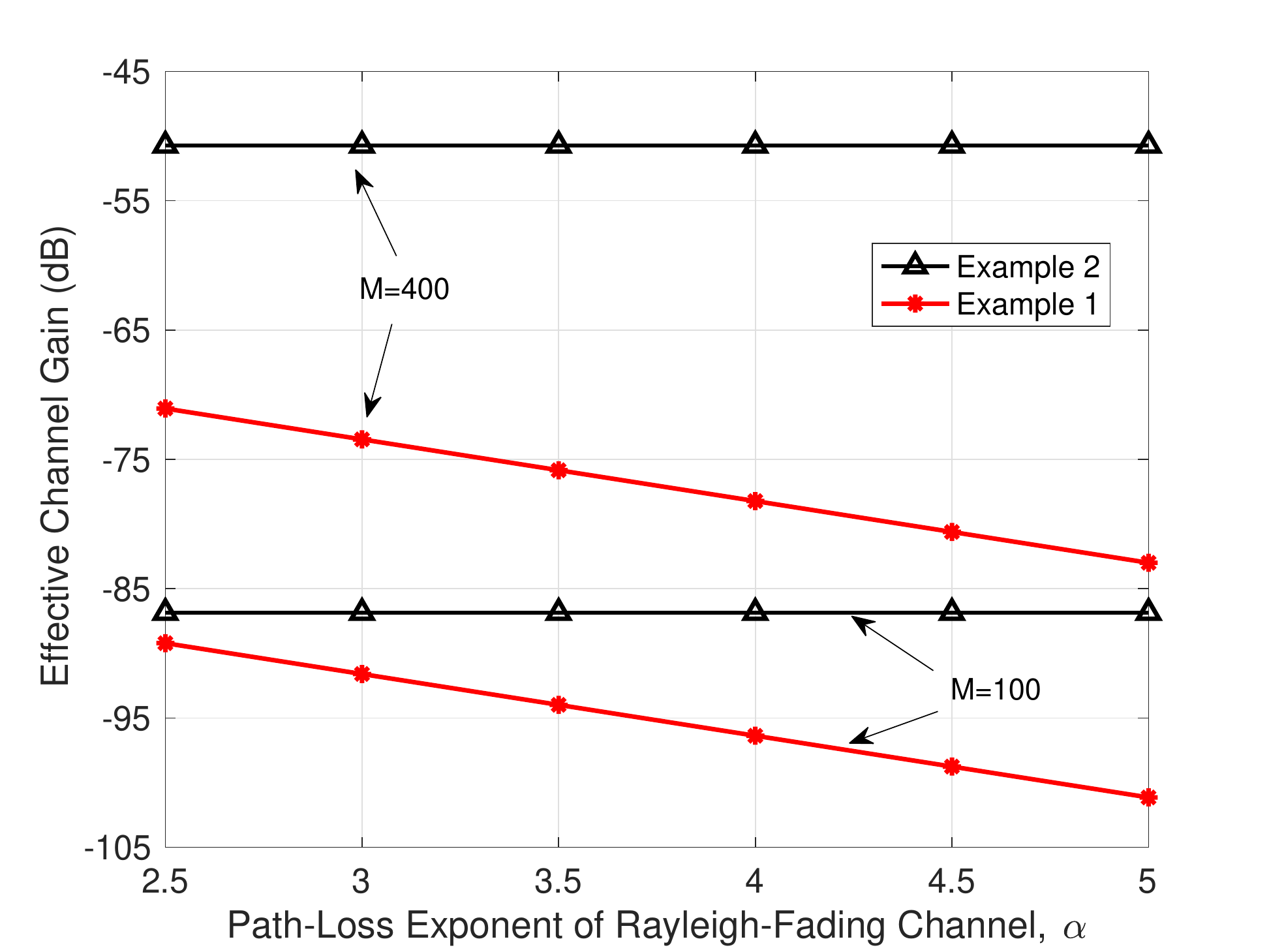}
\caption{Effective channel gain versus path-loss exponent of inter-IRS Rayleigh-fading channels in Example 1, $\alpha$.}\label{NumExp}
\end{minipage}
\vspace{-12pt}
\end{figure}
{\bf{Numerical Example:}} To better manifest the benefits of the proposed multi-IRS aided system, we provide the following two numerical examples, as shown in Fig.\,\ref{NumExpSetup}. In Example 1, as shown in Fig.\,\ref{NumExpSetup}(a), there are two IRSs (labelled as 1 and 2) deployed near the user and BS, respectively, such that LoS-dominant channels can be achieved between the BS and IRS 2 as well as between the user and IRS 1. However, due to the scattered obstacles in the environment, IRS 1 cannot establish an LoS-dominant channel with IRS 2. Accordingly, we assume Rayleigh fading for the channel between them, with a path-loss exponent denoted by $\alpha$. In Example 2, as shown in Fig.\,\ref{NumExpSetup}(b), an additional IRS 3 is properly deployed such that LoS-dominant channels can be established between it and both IRSs 1 and 2. For convenience, in both examples, we assume that each IRS and the BS employ continuous passive and active beamforming, respectively. Moreover, we follow the notations in the previous sections and refer to the BS and user as nodes 0 and 4, respectively. As such, based on (\ref{bmall}), the optimal BS beamforming is given by the MRT ${\mv w}_B(1)={\tilde{\mv h}}_{1,1}/{\lVert {\tilde{\mv h}}_{1,1} \rVert}$ in both examples. Furthermore, in Example 2, based on (\ref{theta2}), the optimal passive beamforming vectors of IRSs 1, 2, and 3 are given by ${\mv \theta}_I(0,1,3)$, ${\mv \theta}_I(1,3,2)$, and ${\mv \theta}_I(3,2,4)$, respectively. However, it is generally difficult to derive the optimal passive beamforming vectors of IRSs 1 and 2 in Example 1 due to the arbitrary-rank channel between them. In this paper, we apply a similar alternating optimization approach as in \cite{zheng2021double} to alternately optimize the passive beamforming of one IRS with that of the other being fixed, until convergence is reached.

In Fig.\,\ref{NumExp}, we plot the effective BS-user channel gain in Example 1 versus the path-loss exponent of the Rayleigh-fading channel between IRSs 1 and 2, $\alpha$, and compare it with that in Example 2. The BS is equipped with $N_B=32$ antennas, while each IRS is equipped with $M=100$ or $400$ reflecting elements. The carrier frequency is set to 5 GHz. All results are averaged over 100 random channel realizations. It is observed from Fig.\,\ref{NumExp} that the effective BS-user channel gain in Example 1 monotonically decreases with $\alpha$ and is lower than that in Example 2. In particular, when $M=400$, it is around 20 dB lower than that in Example 2 even for $\alpha=2.5$. This is expected since no CPB gain can be achieved in Example 1 while a significant CPB gain of $M^6$ can be achieved in Example 2 despite its generally higher end-to-end path loss. Thus, the proposed multi-IRS aided system is practically useful in enhancing the communication link strength in a complex environment with dense obstacles.

\subsection{Problem Formulation}\label{pf}
In this paper, we aim to maximize the minimum signal-to-noise-plus-interference ratio (SINR) achievable by the $K$ users, by optimizing the reflection paths $\Omega^{(k)}, k \in \cal K$, subject to the feasibility constraints in (\ref{feasible1})-(\ref{feasible3}). Due to the well mitigated inter-user interference at each user's receiver, this is equivalent to maximizing the minimum BS-user effective channel gain, i.e., $\mathop {\min}\nolimits_{k \in \cal K} \lvert h_{0,J+k}(\Omega^{(k)})\rvert^2$. The optimization problem is thus formulated as
\begin{equation}\label{op1}
{\text{(P1)}} \mathop {\max}\limits_{\{\Omega^{(k)}\}_{k \in \cal K}}\;\mathop {\min}\limits_{k \in \cal K}\;\; \lvert h_{0,J+k}(\Omega^{(k)}) \rvert^2 \qquad
\text{s.t.}\;\;{\text{(\ref{feasible1})-(\ref{feasible3})}}.
\end{equation}

However, (P1) is a combinatorial optimization problem due to its integer and coupled variables. In addition, $\tilde A^{(k)}_n, k \in \cal K$ in $h_{0,J+k}(\Omega^{(k)})$ are functions of the corresponding AoAs and AoDs if $b_1$ or $b_2$ is finite, while they become a constant $M$ in the case of continuous IRS beamforming, as considered in \cite{mei2020massive} and \cite{mei2020cooperative}. Thus, it is challenging to obtain the optimal solution to (P1) via standard optimization methods in general, especially in the case with a small $b_1$ or $b_2$. To tackle this challenging problem, we reformulate it as an equivalent graph-optimization problem which is then solved, as detailed in the next section. 

\section{Proposed Solution to (P1)}\label{sol}
In this section, we first reformulate (P1) as an equivalent problem in graph theory under the general case with finite $b_1$ and $b_2$, and thereby show that it is NP-complete. Then, a parametrized recursive algorithm is proposed to efficiently solve this problem sub-optimally in general. Finally, we show that (P1) can be more efficiently solved by the proposed algorithm in the special case of continuous IRS beamforming with $b_1$ and $b_2 \rightarrow \infty$.

\subsection{Problem Reformulation via Graph Theory}\label{reform}
Obviously, in (P1), it is equivalent to minimizing the maximum $\lvert h_{0,J+k}(\Omega^{(k)}) \rvert^{-2}$ among all $k \in \cal K$. Based on (\ref{eq1}), we have
\begin{equation}\label{eq4}
	\lvert h_{0,J+k}(\Omega^{(k)}) \rvert^{-2}=\frac{d^2_{0,a^{(k)}_{1}}}{\beta N_B}\cdot\prod\limits_{n=1}^{N_k}\frac{d^2_{a^{(k)}_n,a^{(k)}_{n+1}}}{\beta\lvert \tilde A^{(k)}_n \rvert^2}, k \in {\cal K}.
\end{equation}

Then, by taking the logarithm of (\ref{eq4}), (P1) becomes equivalent to
\begin{equation}\label{op2}
\mathop {\min}\limits_{\{\Omega^{(k)}\}_{k \in \cal K}}\mathop {\max}\limits_{k \in \cal K}\;F(\Omega^{(k)}), \\\quad\text{s.t.}\;\;{\text{(\ref{feasible1})-(\ref{feasible3})}},
\end{equation}
where
\begin{equation}\label{eq5}
	F(\Omega^{(k)})=\ln\frac{d^2_{0,a^{(k)}_{1}}}{\beta N_B}+\sum\limits_{n=1}^{N_k}\ln\frac{d^2_{a^{(k)}_n,a^{(k)}_{n+1}}}{\beta\lvert \tilde A^{(k)}_n \rvert^2}.
\end{equation}

Next, we recast problem (\ref{op2}) as an equivalent problem in graph theory subject to the constraints (\ref{feasible1})-(\ref{feasible3}). Following the similar procedures in \cite{mei2020massive} and \cite{mei2020cooperative}, we construct a directed and unweighted graph $G_0 = (V_0,E_0)$. The vertex set $V_0$ consists of all nodes in the system, i.e., $V_0=\{0,1,2,\cdots,J+K\}$. Furthermore, we consider that each of the $K$ beams can only be routed outwards from one IRS $i$ to a farther IRS $j$ from the BS with $d_{j,0} > d_{i,0}, i,j \in \cal J$, so as to reach its intended user as quickly as possible. Hence, the edge set $E$ is defined as
\begin{align}
E_0=&\{(0,j)| l_{0,j} = 1, j \in {\cal J}\} \nonumber\\
&\cup \{(i,j)|l_{i,j} = 1, d_{j,0} > d_{i,0}, i,j \in {\cal J}\} \nonumber\\
&\cup \{(j,J+k)|\, l_{j,J+k} = 1, j \in {\cal J}, k \in \cal K\},\label{edgeSet}
\end{align}
i.e., there exists an edge from vertex $i$ to vertex $j$ if and only if an LoS path exists between them and $d_{j,0} > d_{i,0}$, except that vertex $j$ corresponds to a user, i.e., $j=J+k, k \in \cal K$. Thus, we have $\lvert E_0 \rvert=\frac{1}{2}\sum\nolimits_{i=0}^{J+K}\sum\nolimits_{j=0}^{J+K}l_{i,j}$. Note that (\ref{edgeSet}) ensures that there is no circle in $G$, i.e., $G$ is a direct acyclic graph (DAG). Given the constructed graph $G$, any reflection path from the BS to user $k$ corresponds to a path from node 0 to node $J+k$ in $G$. However, different from the beam routing problems in \cite{mei2020massive} and \cite{mei2020cooperative} with continuous IRS beamforming with $b_1$ and $b_2 \rightarrow \infty$, it is difficult to assign a weight to each edge in $G_0$ to recast problem (\ref{op2}) as an equivalent graph-optimization problem. This is because each $\tilde A^{(k)}_n$ in (\ref{eq5}) is associated with three vertices, i.e., vertices $a^{(k)}_{n-1}$, $a^{(k)}_n$ and $a^{(k)}_{n+1}$, but each edge in $G_0$ is only associated with two vertices. However, in \cite{mei2020massive} and \cite{mei2020cooperative}, we have $\tilde A^{(k)}_n=M$, which greatly simplifies the weight assignment in $G_0$.

To resolve the above issues, a new DAG of higher dimension, denoted as $G=(V,E)$, should be constructed from $G_0$. Specifically, besides vertex $0$ and vertices $J+k, k \in \cal K$, we create a vertex in $G$ for each edge in $G_0$; while for every two edges in $G_0$ that share a common vertex, we create an edge between their corresponding vertices in $G$. The resulting graph $G$ is known as the {\it line graph} of $G_0$ in graph theory. Mathematically, for $G$, its vertex set $V$ is given by
\begin{equation}\label{vtxAll}
V=\{v_{i,j}|\, (i,j) \in E_0\} \cup \{0,J+1,J+2,\cdots,J+K\}.
\end{equation}
Obviously, we have $\lvert V \rvert = \lvert E_0 \rvert+K+1$. The edge set $E$ is given by
\begin{align}
E=&\{(0,v_{0,j})|\, j \in {\cal J}\} \cup \{(v_{j,J+k},J+k)|\, j \in {\cal J}, k \in {\cal K}\} \nonumber\\
&\cup \{(v_{i,j},v_{j,r})|\, i,j,r \in V\}.\label{edgeAll}
\end{align}
It follows from (\ref{vtxAll}) and (\ref{edgeAll}) that the edge $(0,v_{0,j})$ ($(v_{j,J+k},J+k)$) indicates that there exists an LoS path from the BS (IRS $j$) to IRS $j$ (user $k$). Moreover, the edge $(v_{i,j},v_{j,r})$ indicates that there exist two pairwise LoS paths from node $i$ and node $r$ via IRS $j$. In this new graph $G$, some edges in $E$ involve three vertices, thus making the weight assignment possible. To determine the edge weights in $G$, we first rewrite $F(\Omega^{(k)}), k \in \cal K$ in (\ref{eq5}) as
\begin{equation}\label{eq6}\small
	F(\Omega^{(k)})\!=\!\ln\frac{d_{0,a^{(k)}_1}}{\sqrt\beta N_B}+\ln\frac{d_{a^{(k)}_{N_k},J+k}}{\sqrt\beta}+\sum\limits_{n=1}^{N_k}\ln\frac{d_{a^{(k)}_{n-1},a^{(k)}_n}d_{a^{(k)}_n,a^{(k)}_{n+1}}}{\beta\lvert \tilde A^{(k)}_n \rvert^2},
\end{equation}
by rearranging the terms in it. Accordingly, the weight of each edge in $E$ is set as follows:
\begin{align}
&W(0,v_{0,j})=\ln\frac{d_{0,j}}{\sqrt\beta N_B},\; W(v_{j,J+k},J+k)=\ln\frac{d_{j,J+k}}{\sqrt\beta},\nonumber\\
&W(v_{i,j},v_{j,r})\!\!=\!\!
\begin{cases}\label{weightV}
	\ln\frac{d_{i,j}d_{j,r}}{\beta\lvert {\tilde{\mv s}}^H_{j,r,1}{\rm diag}({\mv \theta}_I(0,j,r)){\tilde{\mv h}}_{j,2}\rvert^2}&{\text{if}}\;i=0\\
	\ln\frac{d_{i,j}d_{j,r}}{\beta\lvert \tilde{\mv g}^H_{j,J+k}{\rm diag}({\mv \theta}_I(i,j,J+k)){\tilde{\mv s}}_{i,j,2} \rvert^2}\!\!\!\!&{\text{if}}\;r=J\!+\!k\\
	\ln\frac{d_{i,j}d_{j,r}}{\beta\lvert {\tilde{\mv s}}^H_{j,r,1}{\rm diag}({\mv \theta}_I(i,j,r)){\tilde{\mv s}}_{i,j,2} \rvert^2}&{\text{otherwise}}.
\end{cases}
\end{align}
Note that the above weights may be negative, e.g., when $M$, $b_1$ and $b_2$ are practically large, such that the argument of the logarithm in (\ref{weightV}) is smaller than one. 

With the constructed line graph $G$, we can establish a one-to-one correspondence between each path from vertex $0$ to vertex $J+k, k \in \cal K$ in $G$ and that in $G_0$. For example, if a path in $G$ is given by $0 \rightarrow v_{0,1} \rightarrow v_{1,3} \rightarrow v_{3,J+1} \rightarrow J+1$, then it corresponds to the path $0 \rightarrow 1 \rightarrow 3 \rightarrow J+1$ in $G_0$ and thus a reflection path from the BS to user 1 via IRSs 1 and 3. In particular, the sum of edge weights of any path from vertex $0$ to vertex $J+k, k \in \cal K$ in $G$ is equal to $F(\Omega^{(k)})$, if its corresponding path in $G_0$ is $\Omega^{(k)}$. Since $G_0$ is a DAG, it is easy to verify that $G$ is also a DAG. Thus, for any path in $G$, its corresponding path in $G_0$ can automatically satisfy the constraints in (\ref{feasible1})-(\ref{feasible2}). To handle the more challenging constraint (\ref{feasible3}), we present the following definitions.
\begin{definition}
	Neighbor-disjoint paths refer to the paths in a graph which do not have any common or neighboring vertices except their starting points.
\end{definition}

According to Definition 1, the constraints in (\ref{feasible3}) can be satisfied if the $K$ paths from vertex $0$ to vertices $J+k, k \in \cal K$ in $G_0$ are neighbor-disjoint. As such, problem (\ref{op2}) is equivalent to the following graph-optimization problem, denoted as (P2).\vspace{6pt}
 \begin{tcolorbox}[standard jigsaw, opacityback=0]
  (P2)\;{\it Find $K$ paths from vertex $0$ to vertices $J+k, k \in \cal K$ in $G$, respectively, such that the length of the longest path (i.e., the path with the maximum sum of edge weights) is minimized and their corresponding paths in $G_0$ are neighbor-disjoint.}
\end{tcolorbox}

Note that neighbor-disjoint routing design has been previously studied in various multi-hop wireless networks, such as ad-hoc networks and wireless sensor networks, for the purpose of load balancing or interference mitigation\cite{teo2008interference,waharte2008probability}. However, most of these works only focus on discovering a set of neighbor-disjoint paths through different medium access control (MAC) layer protocols, but not from an optimal routing design perspective. A common routing design is by utilizing the shortest path algorithm to sequentially update the paths for the $K$ users\cite{teo2008interference}. Specifically, after deriving the shortest path for a user in $G$, the nodes in its corresponding path in $G_0$ (except node 0) and their neighbors are removed. Then, a new line graph $G$ is constructed to determine the shortest path for the next user, so as to satisfy (\ref{feasible3}). However, as will be shown in Section \ref{sim}, this sequential update design generally yields suboptimal paths and even fails to return feasible paths. This is because the set of feasible paths for the current user critically depends on the previously optimized paths for the other users. In fact, it has been proved in \cite{waharte2008probability} that finding $K$ neighbor-disjoint paths in $G_0$ is NP-complete even in the case of $K=2$. As such, (P2) remains a challenging problem, which will be solved next.

\subsection{Proposed Solution to (P2)}\label{propsol}
The basic idea of the proposed solution to (P2) is by first finding $Q\,(\ge 1)$ candidate shortest paths from node 0 to each node $J+k, k \in \cal K$ in $G$ (thus $G_0$). Given these candidate shortest paths, we further construct a new {\it path graph}, based on which a recursive algorithm is performed to partially enumerate the feasible paths and select the best one as the solution to (P2), as specified below.

{\it 1) Step 1: Find the candidate shortest paths.} First, for the nodes 0 and $J+k, k \in \cal K$ in $G$, we invoke the Yen's algorithm\cite{west1996introduction} to find $Q$ candidate shortest paths between them. If the total number of paths between the two nodes is less than $Q$, we assume that there exist additional virtual paths between them with infinite sum of edge weights. For convenience, we denote by $p_k^{(q)}$ and $c_k^{(q)}, k \in {\cal K}, q \le Q$ the $q$-th candidate shortest path between vertices 0 and $J+k$ and its sum of edge weights, respectively. Let ${\cal P}=\{p_k^{(q)}, k \in {\cal K}, q \le Q\}$ be the set of all candidate shortest paths. The time complexity for this step is ${\cal O}(KQ\lvert V \rvert(\lvert E \rvert+\lvert V \rvert\log \lvert V \rvert))$\cite{west1996introduction}.

{\it 2) Step 2: Construct the path graph.} Next, we construct a new undirected graph $G_p=(V_p,E_p)$, where each vertex in $V_p$ corresponds to one candidate shortest path obtained in Step 1 (thus termed as path graph), i.e., $V_p=\{v(p_k^{(q)})\,|\, k \in {\cal K}, q \le Q\}$. Hence, we have $\lvert V_p \rvert = KQ$. By this means, we can establish a one-to-one mapping between any path in $\cal P$ and one vertex in $G_p$. Moreover, since there also exists a one-to-one mapping between any path in $G$ (thus in $\cal P$) and a path in $G_0$, each vertex in $G_p$ also corresponds to a unique path in $G_0$. In particular, the vertex $v(p_k^{(q)})$ in $G_p$ corresponds to a path from vertex 0 to vertex $J+k$ in $G_0$. For example, as shown in Fig.\,\ref{Example}, if the path $0 \rightarrow v_{0,1} \rightarrow v_{1,3} \rightarrow v_{3,J+1} \rightarrow J+1$ is the second candidate shortest path from node 0 to node $J+1$ in $G$ and also included in $\cal P$ (e.g., $Q = 2$), then it corresponds to the vertex $v(p_1^{(2)})$ in $G_p$, which thus corresponds to the path $0 \rightarrow 1 \rightarrow 3 \rightarrow J+1$ in $G_0$. Based on this fact, for any two vertices in $G_p$, we add an edge between them if and only if their corresponding paths in $G_0$ are neighbor-disjoint or satisfy the considered path separation constraints. As $\lvert V_p \rvert = KQ$, we need to execute this procedure $KQ(KQ-1)/2$ times; while in each execution, we need to check the connectivity between any two vertices in the two corresponding paths in $G_0$, respectively, so as to determine whether they are neighbor-disjoint or not. Since the number of vertices in any path in $G_0$ should not exceed $\lvert V_0 \rvert = J+K+1$, the worst-case complexity of this step is given by ${\cal O}(K^2Q^2(J+K)^2)$. Finally, we assign each vertex $v(p_k^{(q)})$ in $G_p$ with a weight, which is equal to the sum of edge weights of its corresponding path in $G$, i.e., $c_k^{(q)}$, obtained in Step 1. 
\begin{figure}[!t]
\centering
\includegraphics[width=3.5in]{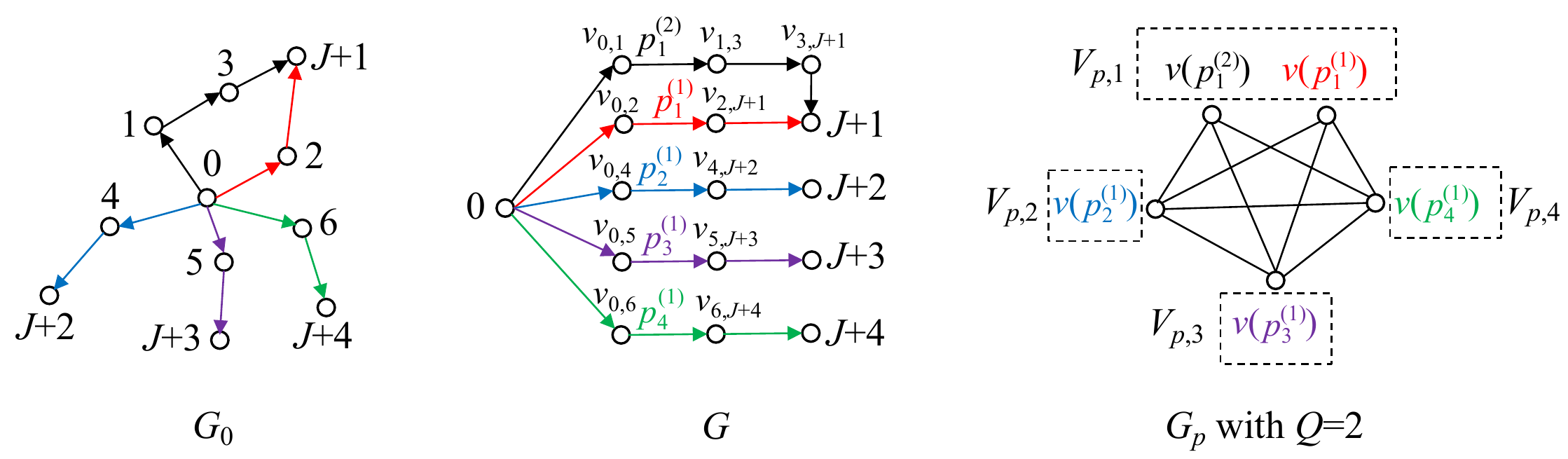}
\DeclareGraphicsExtensions.\vspace{-6pt}
\caption{An example of the constructed graphs with $J=6$ and $K=4$, where the corresponding paths and vertices are marked by the same color.}\label{Example}
\vspace{-9pt}
\end{figure}

To relate the path graph $G_p$ to (P2), we first introduce the following definitions.
\begin{definition}
	A $K$-partite graph refers to a graph whose vertices can be partitioned into $K$ disjoint sets, such that there is no edge between any two vertices within the same set.
\end{definition}
\begin{definition}
	A clique is a subset of vertices of an undirected graph, such that every two distinct vertices in the clique are adjacent. 
\end{definition}

Based on Definitions 2 and 3, we can verify the following facts, which specify the relationship among $G_p$, $G$ and $G_0$.
\begin{fact}
	$G_p$ is a $K$-partite graph, with the $k$-th disjoint set given by $V_{p,k}=\{v(p_k^{(q)}) \,|\,q \le Q\}, k \in \cal K$.
\end{fact}
\begin{fact}
	For $K$ neighbor-disjoint paths from vertex 0 to vertices $J+k, k \in \cal K$ in $G_0$, if their corresponding paths in $G$ are included in $\cal P$, they correspond to a clique of size $K$ in $G_p$. 
\end{fact}

Fact 1 can be proved by noting that for any two vertices in each $V_{p,k}, k \in \cal K$, their corresponding paths in $G_0$ should not be neighbor-disjoint, since they share the same end vertex $J+k$. For Fact 2, it can be easily verified based on Definition 3 and the definition of $G_p$. For example, in Fig.\,\ref{Example}, $G_p$ is a 4-partite graph and consists of two cliques of size 4, i.e., $(v(p_1^{(1)}),v(p_2^{(1)}),v(p_3^{(1)}),v(p_4^{(1)}))$ and $(v(p_1^{(2)}),v(p_2^{(1)}),v(p_3^{(1)}),v(p_4^{(1)}))$, each corresponding to 4 neighbor-disjoint paths in $G_0$. The two vertices $v(p_1^{(1)})$ and $v(p_1^{(2)})$ in $V_{p,1}$ are not connected as their corresponding paths in $G_0$ share the same end vertex $J+1$.

According to Facts 1 and 2, we aim to solve the following clique search problem, denoted as (P3).\vspace{6pt}
 \begin{tcolorbox}[standard jigsaw, opacityback=0]
  (P3)\;{\it Find a clique of size $K$ in a $K$-partite graph $G_p$, whose maximum vertex weight is minimized.}
\end{tcolorbox}
The optimal clique for (P3) corresponds to the best solution to (P2) among the paths in $\cal P$. Thus, if $Q$ is set to be sufficiently large, such that the optimal paths from node 0 to each node $J+k, k \in \cal K$ are included in $\cal P$, the proposed algorithm ensures to find an optimal solution to (P2) (and hence (P1)), if (P3) is optimally solved. Accordingly, by tuning the value of its parameter $Q$, the proposed algorithm can flexibly balance between its performance and complexity.

{\it 3) Step 3: Clique enumeration.} To find the optimal solution to (P3), we can enumerate all cliques of size $K$ in $G_p$ and then compare their respective maximum vertex weights. However, finding all cliques of size $K$ in a graph is also an NP-complete problem in general when $K > 2$\cite{west1996introduction}. As such, we propose a recursive algorithm to achieve this purpose by leveraging the $K$-partite property of $G_p$, thereby optimally solving (P3). 

Specifically, we will show that each clique of size $K$ in $G_p$ can be recursively constructed based on the cliques of smaller sizes. Note that its $K$ vertices must be selected from the $K$ disjoint sets $V_{p,k}, k \in \cal K$, respectively. Without loss of optimality, we assume that its $k$-th vertex is selected from $V_{p,k}$. Accordingly, let $\Omega_r, r \le K$ denote the set of all cliques of size $r$ in $G_p$, with the $s$-th vertex of each clique in $\Omega_r$ selected from $V_{p,s}, s=1,2,\cdots,r$. Obviously, we have $\Omega_1=V_{p,1}$. Moreover, for each clique (of size $r$) in $\Omega_r, r \le K-1$, if there exists a vertex in $V_{p,r+1}$ which is adjacent to all vertices in this clique, then a new clique (of size $r+1$) in $\Omega_{r+1}$ can be constructed by appending the vertex to this clique. As such, based on the initial condition for $\Omega_1$ and the recursion for $\Omega_r, r \le K-1$, all cliques of size $K$ in $G_p$ can be enumerated in the set $\Omega_K$, which requires the worst-case complexity of ${\cal O}(Q^K)$. To further reduce complexity, it is noted that when a clique of size $K\!-\!1$ is constructed, among all feasible vertices in $V_{p,K}$, we only need to append the vertex with the lowest weight to it. This is because the cliques obtained by appending other feasible vertices cannot yield a lower maximum vertex weight. Thus, the worst-case complexity of the above recursive algorithm can be reduced to ${\cal O}(Q^{K-1})$. In fact, since the number of feasible vertices may significantly decrease when increasingly larger cliques are constructed (owing to the more stringent adjacency constraint), the actual complexity of the proposed recursive enumeration is much lower than ${\cal O}(Q^{K-1})$, as will be shown in Section \ref{sim}. 

Denote by ${\cal C}_i$ the $i$-th clique (of size $K$) in $\Omega_K$ after the enumeration. For each ${\cal C}_i \in \Omega_K$, we can obtain the maximum vertex weight among all of its $K$ vertices, denoted as \[c_i = \mathop {\max}\limits_{v(p_k^{(q)}) \in {\cal C}_i} c_k^{(q)}.\] Thus, the best clique in $\Omega_K$ can be obtained as ${\cal C}_{i^\star}$, with $i^\star \triangleq \arg \mathop {\min}\limits_{i} c_i$. The main procedures of the proposed clique enumeration method for solving (P3) are summarized in Algorithm \ref{Alg1}, where a function ``$\textsc{RecEnum}$'' is defined and recursively called to achieve the recursive enumeration. 
\begin{algorithm}
  \caption{Proposed Clique Enumeration Method for Solving (P3)}\label{Alg1}
  \begin{algorithmic}[1]
  \State Initiate $r=1$ and a clique ${\cal C}=\emptyset$.
  \State Execute $\textsc{RecEnum\,}(r,{\cal C})$ and obtain $\Omega_K$.
  \State Compare the maximum vertex weights for all obtained cliques in $\Omega_K$, i.e., $c_i$'s, and determine the best clique ${\cal C}_{i^\star}$.
  \Function{RecEnum\,}{$r,{\cal C}$}
    \If {$r=K$}
    \State \begin{varwidth}[t]{0.87\linewidth}Among all vertices in $V_{p,K}$ which are adjacent to every vertex in ${\cal C}$, append the vertex with the lowest weight to ${\cal C}$ and obtain a new clique of size $K$, ${\cal C}'$.\end{varwidth}\vspace{3pt}
    \State Add ${\cal C}'$ to the set $\Omega_K$.
    \Else
    \State Initialize $s=1$.
    \While {$s \le Q$}
    \If {${\cal C}=\emptyset$ or the $s$-th vertex in $V_{p,r}$ is adjacent\par\hspace{0.75cm} to every vertex in ${\cal C}$}
    \State \begin{varwidth}[t]{0.74\linewidth}Append this vertex to ${\cal C}$ and obtain a clique of size $r$, ${\cal C}'$.\end{varwidth}\vspace{3pt}
    \State \begin{varwidth}[t]{0.75\linewidth}Add ${\cal C}'$ to the set $\Omega_r$ and execute $\textsc{RecEnum\,}(r+1,{\cal C}')$.\end{varwidth}\vspace{3pt}
    \EndIf
    \State Update $s=s+1$.
    \EndWhile
    \EndIf
  \EndFunction
  \end{algorithmic}
\end{algorithm}

{\it 4) Step 4: Map and output.} Finally, a generally suboptimal MBMH routing solution with a finite value of $Q$ can be obtained by mapping the $K$ vertices in ${\cal C}_{i^\star}$ to $K$ neighbor-disjoint paths in $G_0$. The process of solving (P2) is summarized in Algorithm \ref{Alg2}. The worst-case complexity of Algorithm \ref{Alg2} is given by the sum of the complexity of the first three steps, i.e., ${\cal O}(KQ\lvert V \rvert\lvert E \rvert+KQ\lvert V \rvert^2\log \lvert V \rvert+K^2Q^2(J+K)^2+Q^{K-1})$. Since the active beam search in (\ref{search3}) and the passive beam search in (\ref{bs1}) and (\ref{bs2}) with $r \in \cal J$ can be conducted offline, the online complexity of the proposed MBMH routing design is given by the sum of the complexity of Algorithm 2 and that of the passive beam search for the $K$ users at their associated final-hop IRSs.

It is worth noting that if $\Omega_K = \emptyset$ with a given $Q$ after performing Algorithm \ref{Alg1}, this indicates that (P3) is infeasible. To obtain a feasible clique of size $K$, the value of $Q$ can be increased to enlarge the solution set of (P3). However, if (P3) is still infeasible even with the maximum allowable $Q$, then it can be claimed that (P2) (thus (P1)) is infeasible. As such, some users would be denied access to the considered system. This may happen if the number of users is large (e.g., $K > J$) or some users are close to each other, such that the path separation constraints cannot be met. Besides, if the number of IRSs is small or they are not properly deployed, (P2) may also be infeasible due to the limited number of reflection paths. In this case, the proposed algorithms can help determine the optimal user selection and the reflection paths for the selected users. Specifically, let $K^\prime$ be the maximum number of users that can be granted access to the considered system, which is given by the maximum value of $k$ such that $\Omega_k \ne \emptyset$. Then, the selected users and their reflection paths can be obtained by mapping the best clique (of size $K^\prime$) in $\Omega_{K^\prime}$ to $K^\prime$ neighbor-disjoint paths in $G_0$.
\begin{algorithm}
  \caption{Proposed Algorithm for Solving (P2)}\label{Alg2}
  \begin{algorithmic}[1]
    \State Input the line graph $G$ and the number of candidate shortest paths for each user, $Q$.
    \State Find $Q$ candidate shortest paths from vertex 0 to each vertex $J+k, k \in \cal K$ by invoking the Yen's algorithm and determine the path set ${\cal P}=\{p_k^{(q)}, k \in {\cal K}, q \le Q\}$.
    \State Construct the path graph $G_p=(V_p,E_p)$ with the following steps based on $\cal P$.
    \State \quad {\bf a)} \begin{varwidth}[t]{0.95\linewidth} Make a vertex for each path in $\cal P$, i.e., $V_p=\{v(p_k^{(q)})\,|\, k \in {\cal K}, q \le Q\}$.\end{varwidth}\vspace{3pt}
    \State \quad {\bf b)} \begin{varwidth}[t]{0.9\linewidth} Add an edge between any two vertices in $G_p$ if their corresponding paths in $G_0$ are neighbor-disjoint to determine $E_p$.\end{varwidth}
    \State \quad {\bf c)} \begin{varwidth}[t]{0.9\linewidth} Assign each vertex $v(p_k^{(q)})$ in $G_p$ with the weight $c_k^{(q)}$.\end{varwidth}\vspace{3pt}
    \State Obtain ${\cal C}_{i^\star}$ by performing Algorithm \ref{Alg1}.
    \State Map the $K$ vertices in ${\cal C}_{i^\star}$ to $K$ neighbor-disjoint paths in $G_0$ and output them.
  \end{algorithmic}
\end{algorithm}

\subsection{Special Case with Continuous IRS Beamforming}
If the continuous beamforming with $b_1$ and $b_2 \rightarrow \infty$ is applied at each IRS, (P1) can be more efficiently solved based on $G_0$, without the need of constructing its line graph $G$. Specifically, since we have ${\tilde A}_n^{(k)} = M$ in this case, (\ref{eq4}) becomes
\begin{equation}\label{eq7}
	\lvert h_{0,J+k}(\Omega^{(k)}) \rvert^{-2}=\frac{M^2}{N_B}\prod\limits_{n=0}^{N_k}\frac{d^2_{a^{(k)}_n,a^{(k)}_{n+1}}}{M^2\beta}, k \in {\cal K}.
\end{equation}

By taking the logarithm of (\ref{eq7}) and discarding irrelevant constant terms therein, (P1) becomes equivalent to
\begin{equation}\label{op3}
\mathop {\min}\limits_{\{\Omega^{(k)}\}_{k \in \cal K}}\mathop {\max}\limits_{k \in \cal K}\;\sum\limits_{n=0}^{N_k}\ln\frac{d_{a^{(k)}_n,a^{(k)}_{n+1}}}{M\sqrt\beta}, \\\quad\text{s.t.}\;\;{\text{(\ref{feasible1})-(\ref{feasible3})}}.
\end{equation}

Similarly as in Section \ref{reform}, we construct the DAG $G_0 = (V_0,E_0)$. However, according to (\ref{op3}), we can directly assign a weight to each edge $(i,j)$ in $E_0$, denoted as $W(i,j)=\ln\frac{d_{i,j}}{M\sqrt\beta}$. As a result, (P2) reduces to finding $K$ neighbor-disjoint paths from vertex $0$ to vertices $J+k, k \in \cal K$ in $G_0$, respectively, such that the length of the longest path is minimized. To solve this simplified problem, Algorithm \ref{Alg2} can be similarly applied. The only difference is that the input graph is $G_0$ instead of $G$.

\begin{figure}[htbp]
\centering
\centering
\subfigure[3D plot]{\includegraphics[width=0.42\textwidth]{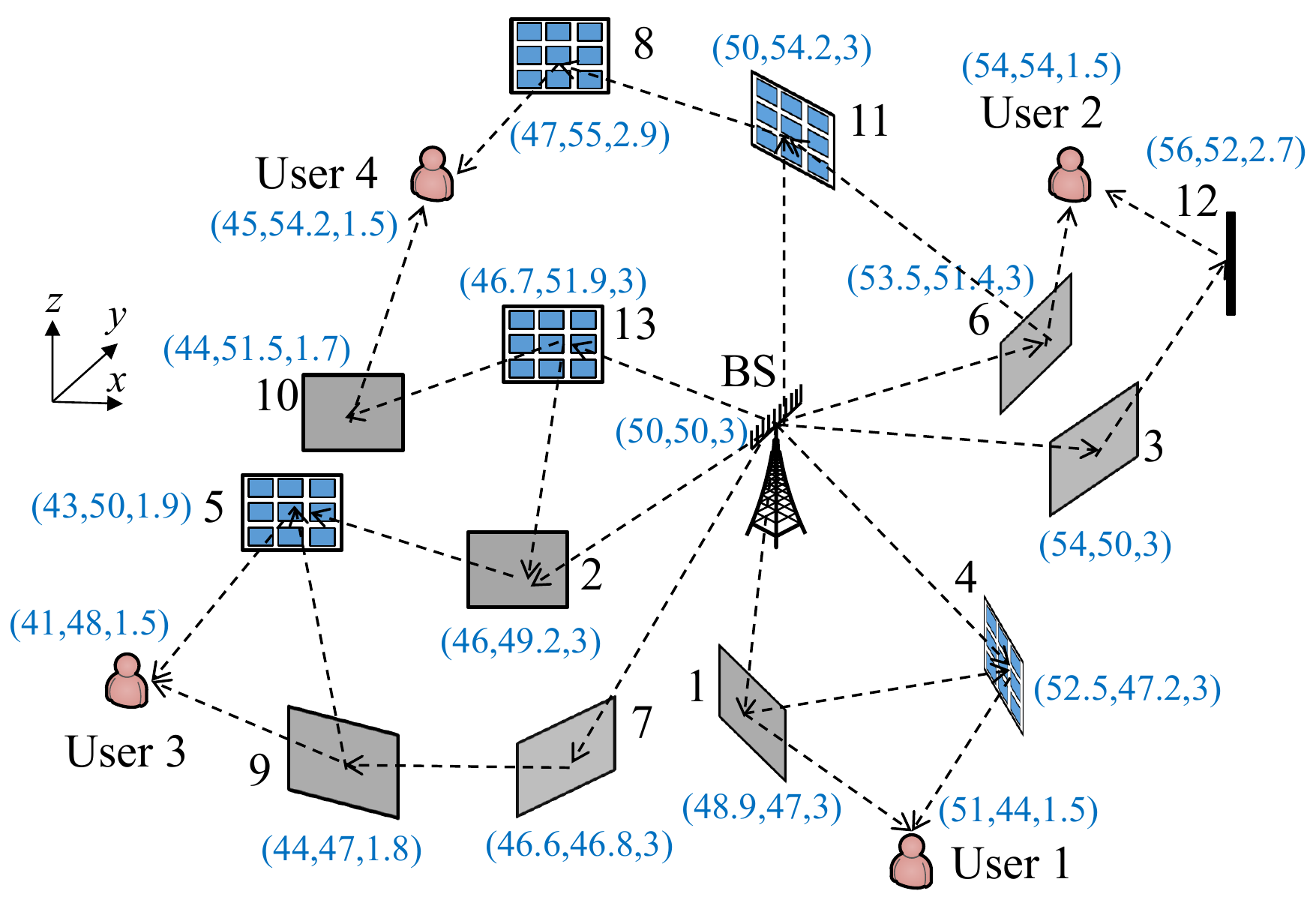}}
\subfigure[Graph representation]{\includegraphics[width=0.46\textwidth]{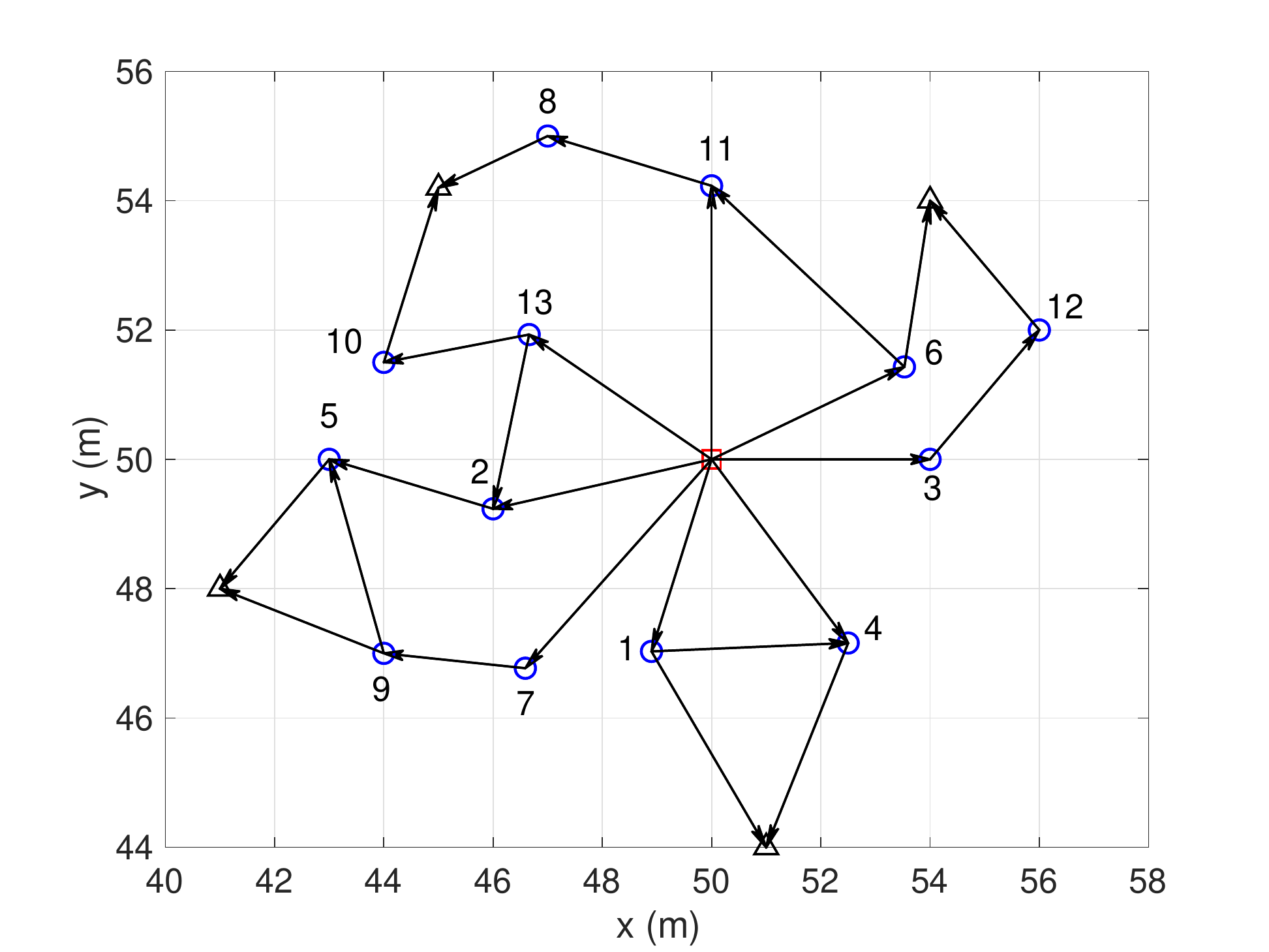}}
\caption{Simulation setup.}\label{topology}
\vspace{-12pt}
\end{figure}
\section{Numerical Results}\label{sim}
In this section, we provide numerical results to evaluate our proposed MBMH routing design. We focus on an indoor multi-IRS aided system (e.g., in a smart factory) with $K=4$ users and $J=13$ IRSs. The 3D coordinates of all nodes, their available LoS links, as well as the facing directions of all IRSs are shown in Fig.\,\ref{topology}(a). The system is assumed to operate at a carrier frequency of 5 GHz. Thus, the carrier wavelength is $\lambda=0.06$ m and the LoS path gain at the reference distance 1 m is $\beta=(\lambda/4\pi)^2=-46.4$ dB. Based on the LoS probability specified in \cite{3GPP38901}, for any two nodes $i$ and $j$ that satisfy the facing condition, we consider that there is an LoS-dominant link between them, i.e., $l_{i,j}=1, i,j \in V$, if its occurrence probability is equal to one, or $d_{i,j} \le$ 5 m. Whereas if $l_{i,j}=0$, we assume that there exist rich scatterers between nodes $i$ and $j$ and model their channel as Rayleigh fading with a path-loss exponent of $\alpha$. The antenna and element spacing at the BS and each IRS are set to $d_A=\lambda/2$ and $d_I=\lambda/4$, respectively. Moreover, we set the minimum distance for far-field propagation as $d_0=$ 2.5 m. Accordingly, the graph representation of the considered multi-IRS aided system, i.e., $G_0$, is shown in Fig.\,\ref{topology}(b). The numbers of elements in each IRS's horizontal and vertical dimensions are set to be identical as $M_0 \triangleq \sqrt{M}=M_1=M_2$. The BS is equipped with $N_B=32$ antennas. We use the $N_B$-point DFT-based codebook as the BS's codebook ${\cal W}_B$, which equally divides the spatial domain $[0,2)$ into $N_B$ sectors. Specifically, let ${\mv w}_{B,i} \in {\mathbb C}^{N_B \times 1}$ denote the $i$-th beam pattern in ${\cal W}_B$. We have
\begin{equation}\label{dft1}
{\mv w}_{B,i}=\frac{1}{\sqrt{N_B}}{\mv e}\Big(\frac{2(i-1)}{N_B},N_B\Big), i=1,2,\cdots,N_B.
\end{equation}
It is verified via simulation that with the deployment of IRSs in Fig.\,\ref{topology} and the codebook in (\ref{dft1}), the asymptotically favorable propagation in (\ref{eq2}) can be achieved for all links between the BS (node 0) and the possible first-hop IRSs (the neighbors of node 0 in $G_0$). The numbers of controlling bits for each IRS's codebooks in the horizontal and vertical dimensions, i.e., ${\cal W}^{(1)}_{I}$ and ${\cal W}^{(2)}_{I}$, are assumed to be identical as $b_0 \triangleq b/2 = b_1=b_2$. Thus, the number of beam patterns in ${\cal W}^{(1)}_{I}$ and ${\cal W}^{(2)}_{I}$ is identical to $D_0 \triangleq 2^{b_0} = \sqrt{D}$. Similarly as the BS, the $D_0$-point DFT codebook is used for ${\cal W}^{(1)}_{I}$ and ${\cal W}^{(2)}_{I}$, while the proposed MBMH routing design is applicable to any IRS beamforming codebook. Let $\mv \theta^{(1)}_{I,i}$ and $\mv \theta^{(2)}_{I,i}$ denote the $i$-th beam patterns in ${\cal W}^{(1)}_{I}$ and ${\cal W}^{(2)}_{I}$, respectively. Then, we have
\begin{equation}\label{dft2}
\mv \theta^{(1)}_{I,i}=\mv \theta^{(2)}_{I,i}={\mv e}\Big(\frac{2(i-1)}{D_0},M_0\Big), i=1,2,\cdots,D_0.
\end{equation}
In the proposed recursive algorithm, the number of candidate shortest paths for each node $J+k$ or user $k, k \in \cal K$ is set to $Q=5$. 

\begin{figure*}[htbp]
\centering
\subfigure[$M_0=24, b_0=7$, without (\ref{feasible3})]{\includegraphics[width=0.48\textwidth]{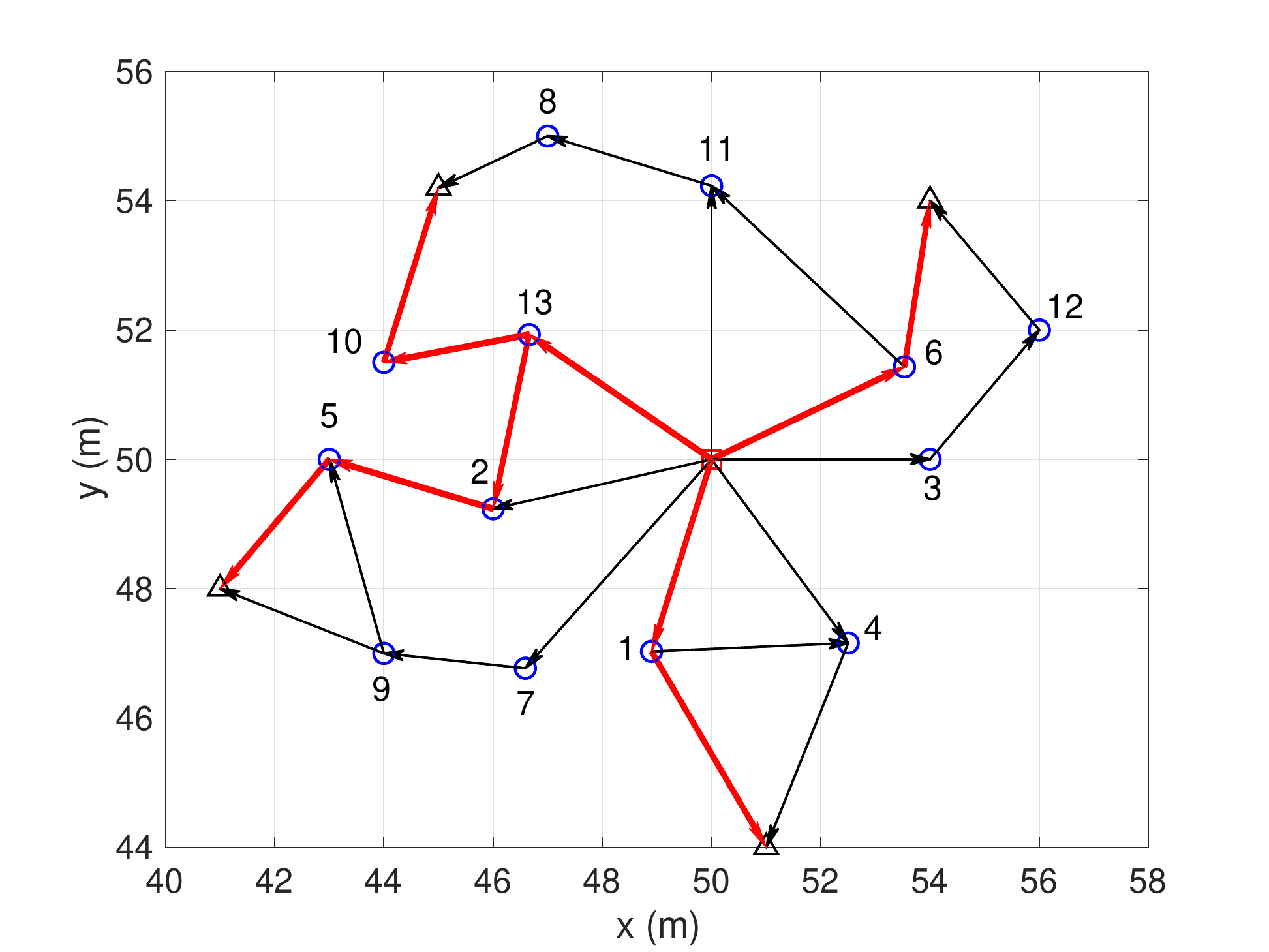}}
\subfigure[$M_0=24, b_0=7$, with (\ref{feasible3})]{\includegraphics[width=0.48\textwidth]{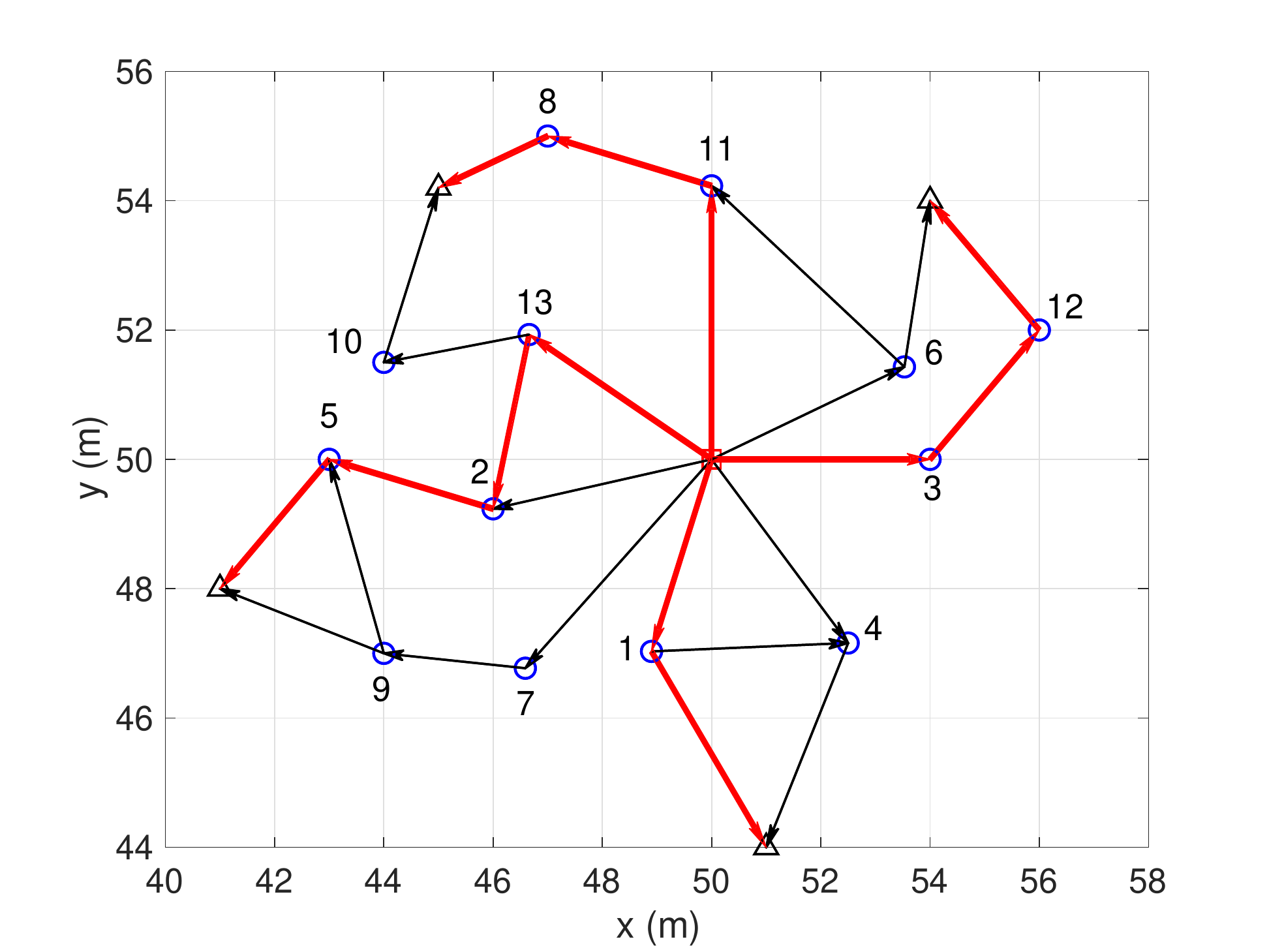}}
\subfigure[$M_0=24, b_0=5$, with (\ref{feasible3})]{\includegraphics[width=0.48\textwidth]{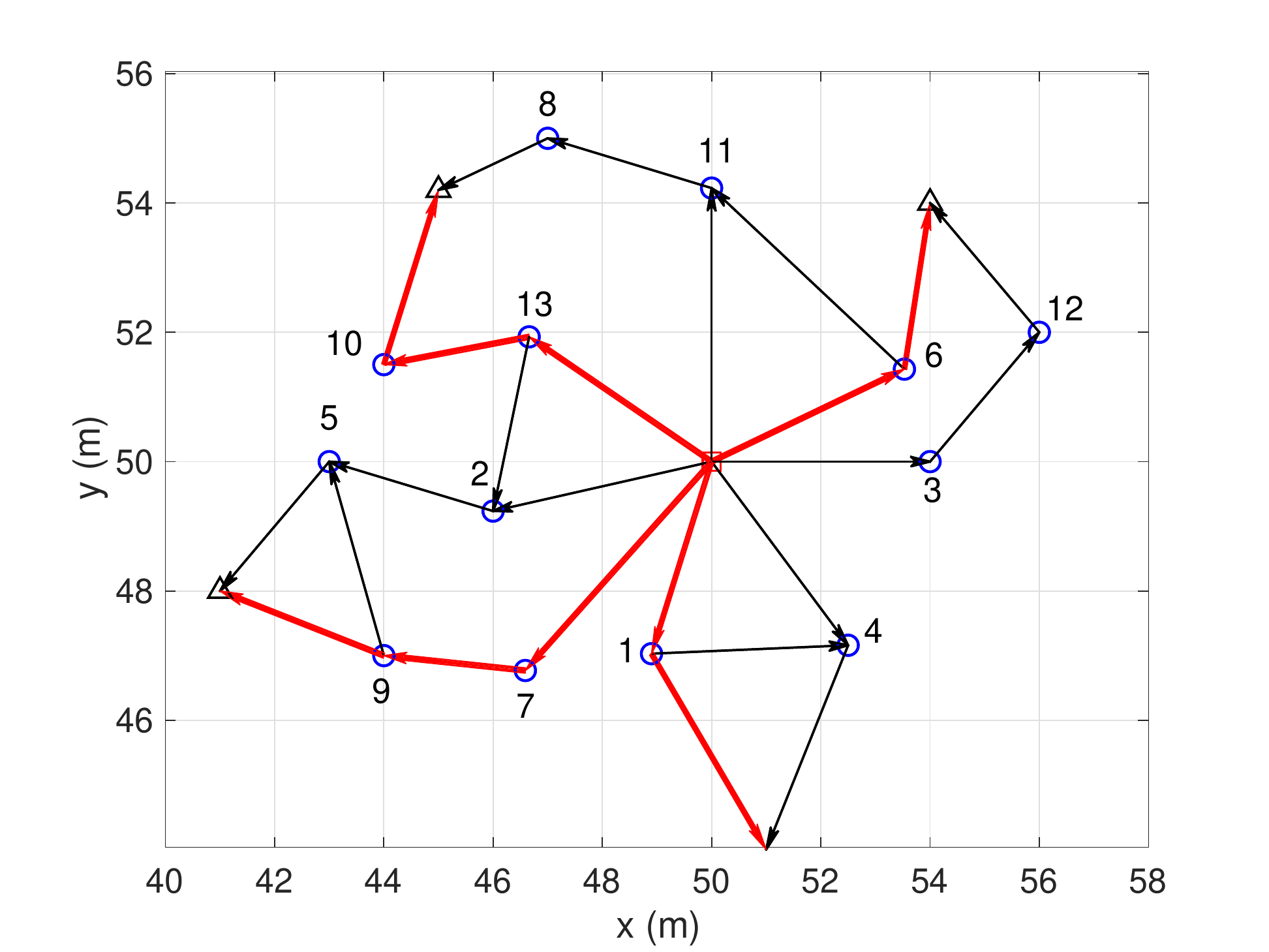}}
\subfigure[$M_0=28, b_0=7$, with (\ref{feasible3})]{\includegraphics[width=0.48\textwidth]{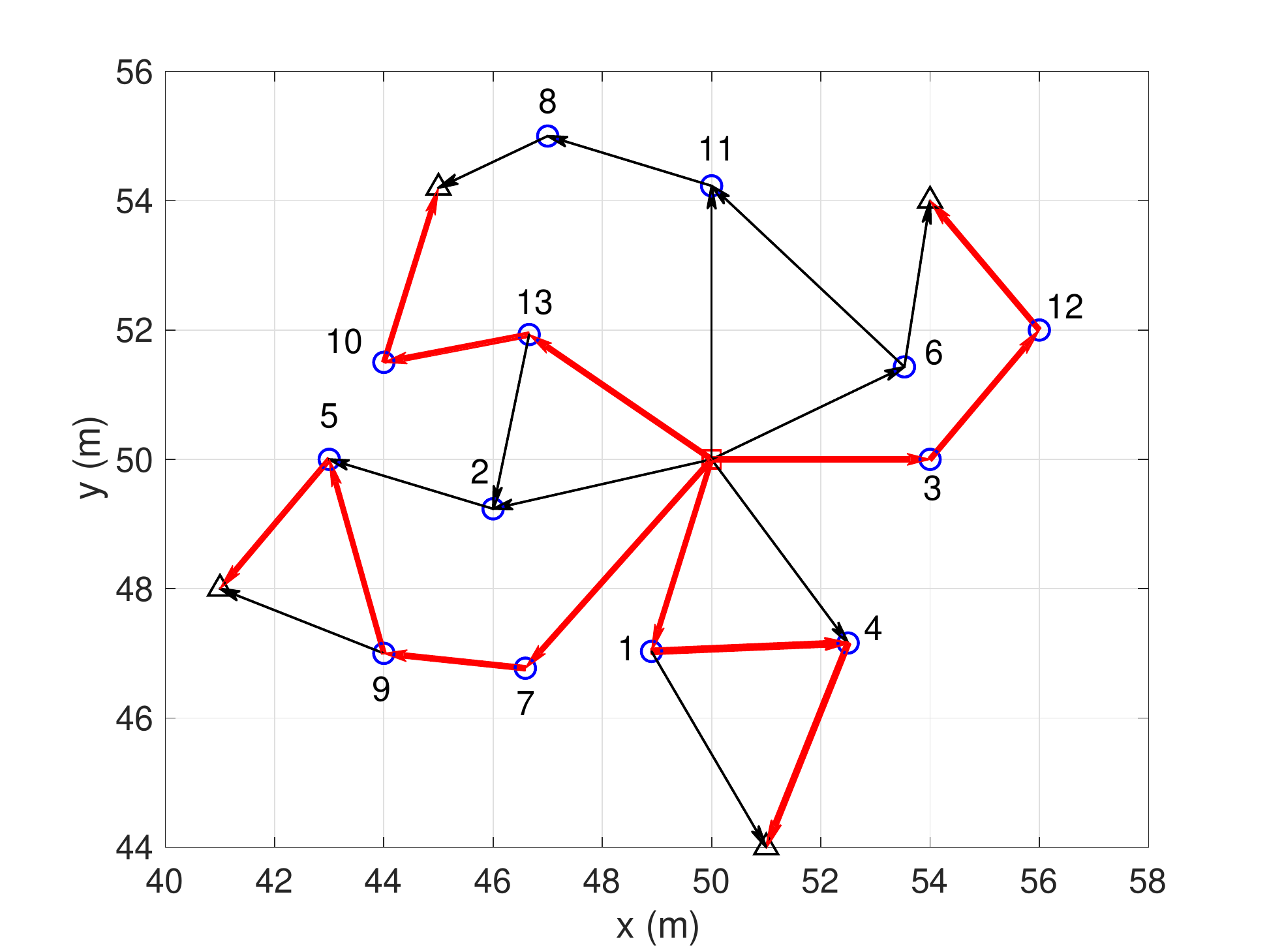}}
\caption{Optimized reflection paths under different setups.}\label{optroute}
\vspace{-6pt}
\end{figure*}
First, Fig.\,\ref{optroute} shows the optimized reflection paths for all users under different numbers of IRS reflecting elements and controlling bits for the IRS codebook in each dimension, i.e., $M_0$ and $b_0$. In Fig.\,\ref{optroute}(a), by utilizing the Bellman-Ford algorithm\cite{west1996introduction} for the shortest path problem on $G$, we plot the optimal reflection path for each user without the path separation constraints in (\ref{feasible3}) under $M_0=24$ (i.e., $M=576$) and $b_0=7$ (i.e., $b=14$) bits. It is observed that IRS 13 exists in the reflection paths for both users 3 and 4. Thus, the proposed algorithm is needed to obtain a feasible MBMH routing solution to (P1) that meets (\ref{feasible3}). In Figs.\,\ref{optroute}(b)-\,\ref{optroute}(d), we plot the optimized MBMH routing solutions by the proposed algorithm subject to (\ref{feasible3}). By comparing Fig.\,\ref{optroute}(b) with Fig.\,\ref{optroute}(a), it is observed that the paths for users 2 and 4 are changed due to the path separation constraints in (\ref{feasible3}). As a result, their effective channel gains with the BS are sacrificed in order to yield $K=4$ neighbor-disjoint paths between the BS and all users. On the other hand, by comparing Fig.\,\ref{optroute}(b) with Fig.\,\ref{optroute}(c), it is observed that for a given $M_0$, increasing the resolution of the IRS codebook may lead to different optimized paths. This is expected since with a larger $b_0$, each IRS has a higher degree of freedom in controlling the direction of the reflected signal, which may result in different reflection paths. Next, by comparing Fig.\,\ref{optroute}(b) with Fig.\,\ref{optroute}(d), it is observed that when $b_0=7$, the optimized paths for some users, e.g., user 1, may go through more IRSs under $M_0=28$ than those under $M_0=24$. This is due to the different dominating effects of the end-to-end path loss and the CPB gain in maximizing the users' effective channel gains with the BS as $b_0$ becomes large. In particular, as $M_0=24$, minimizing the end-to-end path loss is dominant over maximizing the CPB gain. However, as $M_0$ increases to $28$, maximizing the CPB gain becomes more dominant. Since the CPB gain monotonically increases with the hop count of the reflection path when $b_0$ is large, the optimized reflection paths generally go through more IRSs. 

\begin{figure}
\centering
\begin{minipage}[t]{0.49\textwidth}
\centering
\includegraphics[width=3.4in]{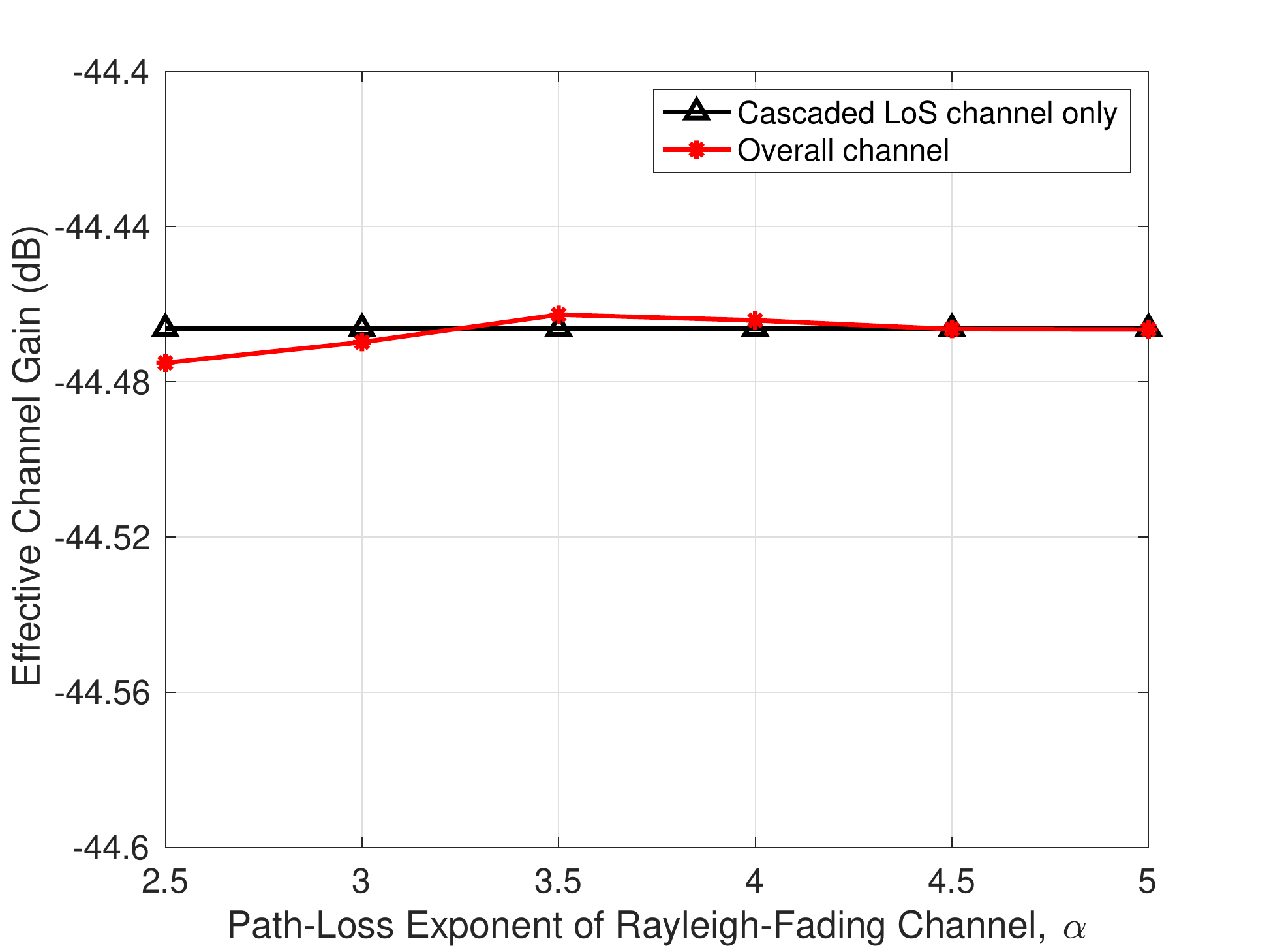}
\caption{Overall channel gain and cascaded LoS channel power versus path-loss exponent of Rayleigh-fading channels, $\alpha$.}\label{DesSigvsPL}
\end{minipage}
\hfill
\begin{minipage}[t]{0.49\textwidth}
\centering
\includegraphics[width=3.4in]{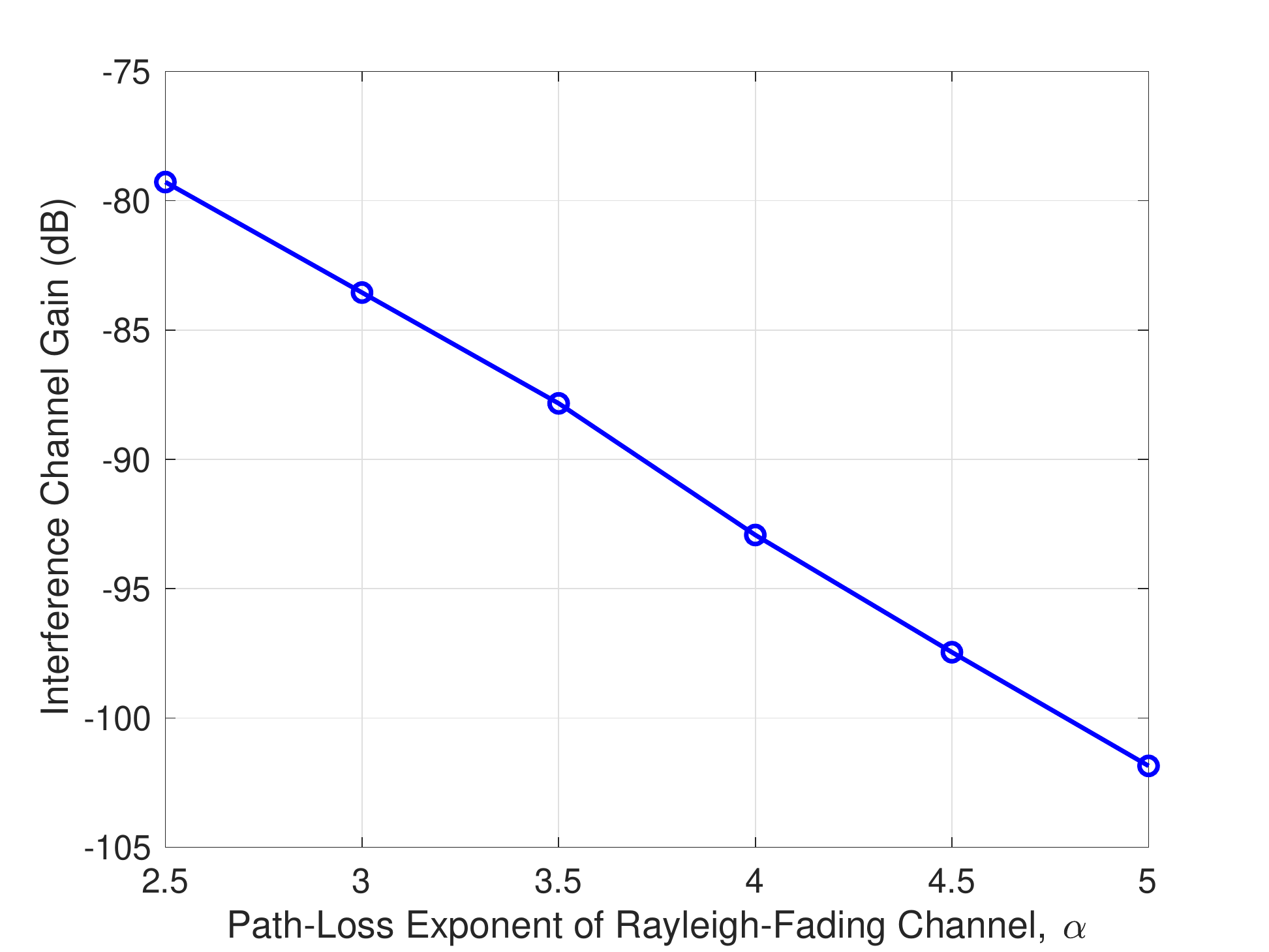}
\caption{Total interference channel gain versus path-loss exponent of inter-IRS Rayleigh-fading channels, $\alpha$.}\label{IntSigvsPL}
\end{minipage}
\vspace{-12pt}
\end{figure}
To measure the strength of the scattered links in the system, under the optimized MBMH routing solution in Fig.\,\ref{optroute}(b), we plot in Fig.\,\ref{DesSigvsPL} the power of the overall channel between the BS and user 4, i.e., the cascaded LoS channel via IRSs 8 and 11 (or $\Omega^{(4)}$) plus other scattered channels by active IRSs in the system, versus the path-loss exponent of Rayleigh-fading channels, $\alpha$. As there are two hops in $\Omega^{(4)}$, we only consider all single-hop and double-hop scattered links between the BS and user 4, due to the more severe multiplicative path-loss and lack of CPB gain over other multi-hop scattered links. The results are averaged over 100 channel realizations. It is observed from Fig.\,\ref{DesSigvsPL} that the overall channel gain is comparable to the cascaded LoS channel gain over the whole range of $\alpha$ considered. This indicates that the strength of the scattered links in the system is practically much lower than that of $\Omega^{(4)}$ and thus can be ignored. Furthermore, to evaluate the performance of the proposed algorithm in terms of mitigating the inter-path interference, we plot in Fig.\,\ref{IntSigvsPL} the total interference channel gain between the BS and user 4 for serving the other three users. By comparing Fig.\,\ref{IntSigvsPL} with Fig.\,\ref{DesSigvsPL}, it is observed that even with $\alpha=2.5$, the strength of the interfering links is about 35.5 dB lower than that of $\Omega^{(4)}$. As $\alpha$ increases, the former further decreases and becomes around 57.5 dB lower than the latter with $\alpha=5$. Given a common signal-to-noise ratio (SNR) level between 20 and 30 dB in modern data transmission, the inter-user interference has been well suppressed below the receiver noise (including other non-IRS reflected interference) in the considered system.

\begin{figure}[htbp]
\centering
\centering
\subfigure[$b_0=5$]{\includegraphics[width=0.48\textwidth]{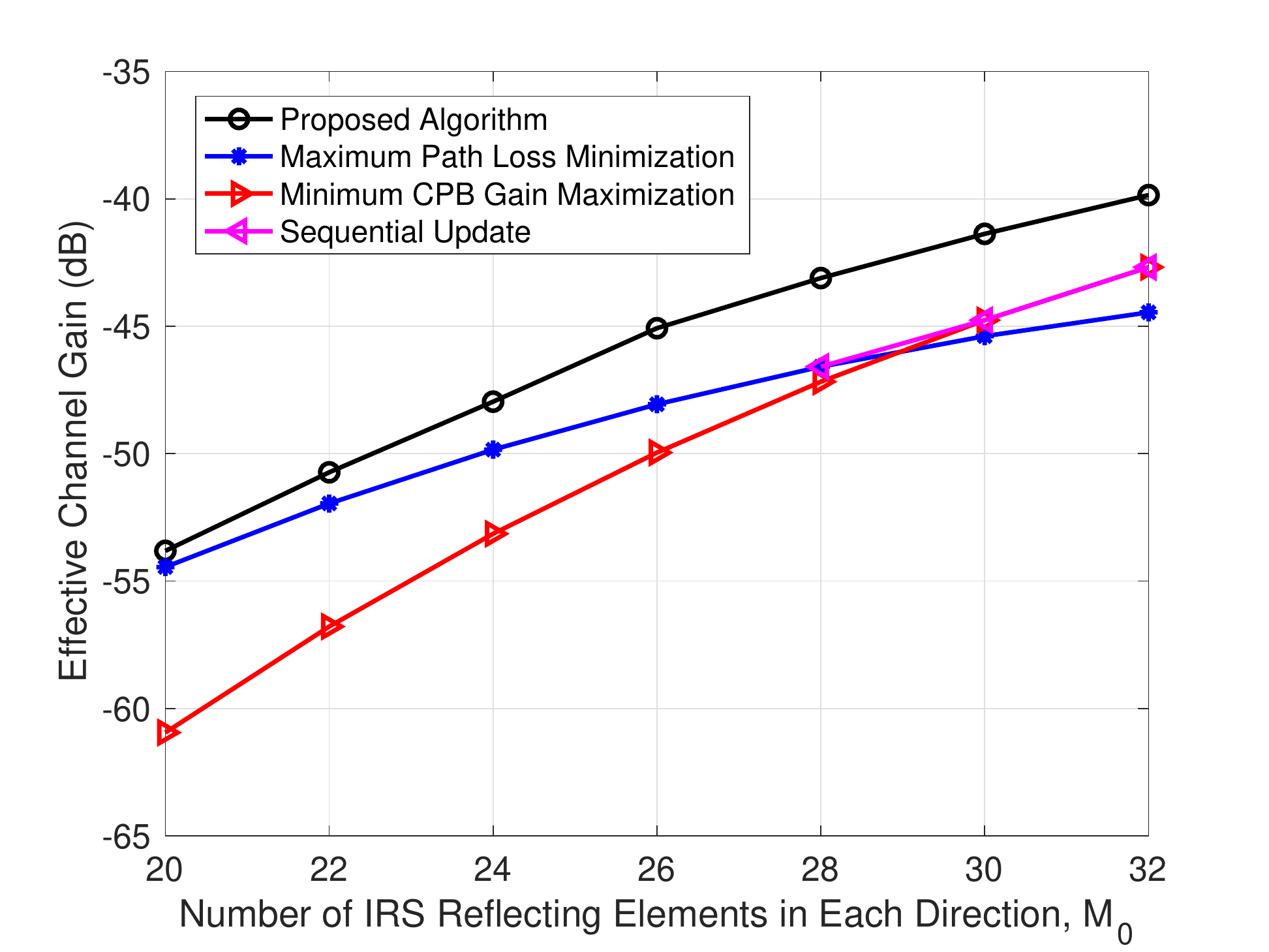}}
\subfigure[$b_0=7$]{\includegraphics[width=0.48\textwidth]{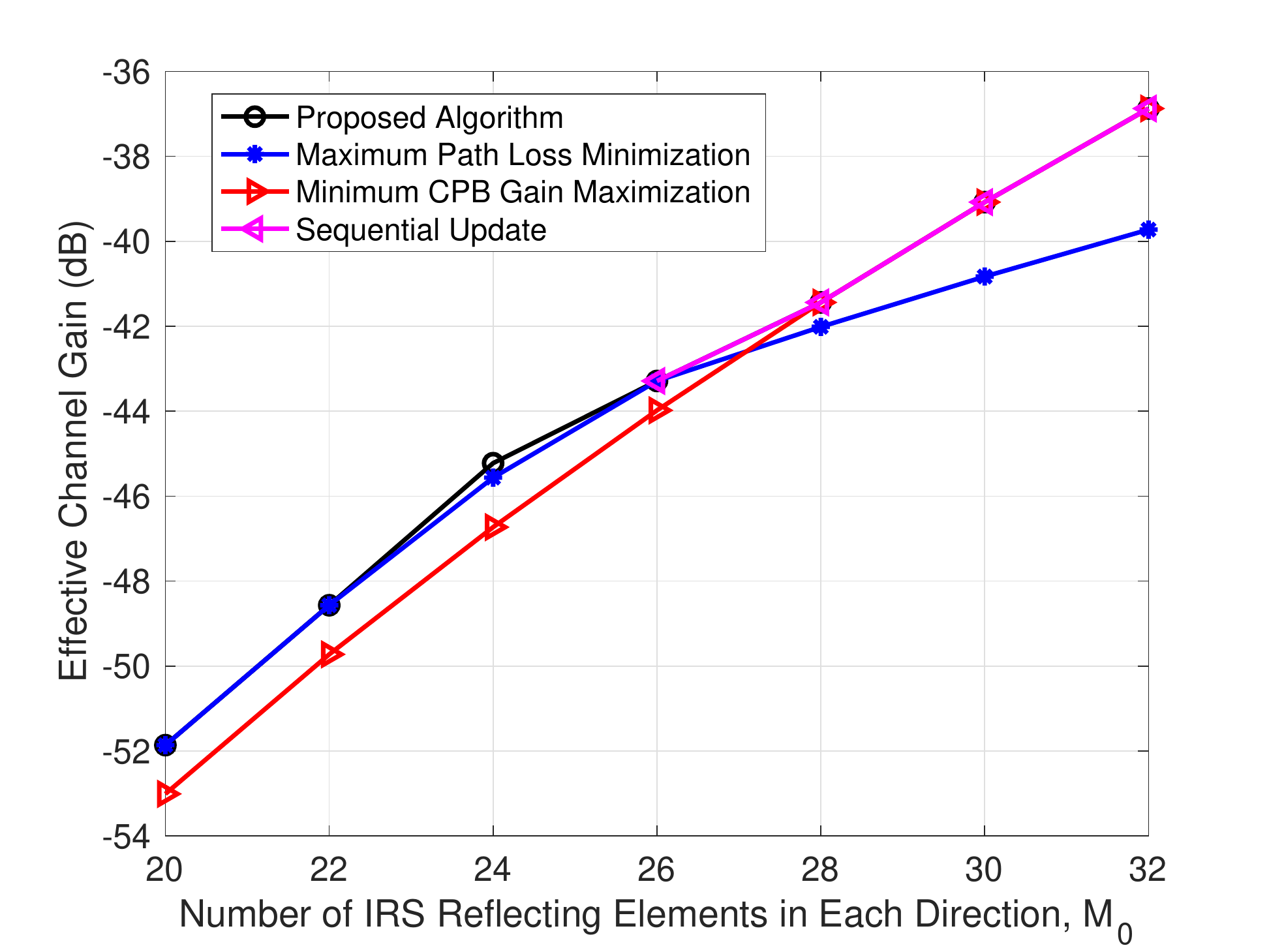}}
\caption{Max-min channel gain versus number of IRS reflecting elements in each dimension, $M_0$.}\label{ChPwvsM}
\vspace{-6pt}
\end{figure}
Next, Fig.\,\ref{ChPwvsM} shows the max-min BS-user channel gain among all users by different schemes versus the number of IRS reflecting elements in each dimension, $M_0$, under $b_0=5$ and $b_0=6$. For performance comparison, we consider the following three benchmark schemes. The {\it first} benchmark is the sequential update scheme, as mentioned at the end of Section \ref{reform}, where we sequentially update the reflection paths from user 1 to user 4. The {\it second} benchmark minimizes the maximum path loss among all BS-user LoS links, while the {\it third} benchmark maximizes the minimum CPB gain among all BS-user LoS links. Their corresponding reflection paths can be obtained by assuming unit CPB gain and unit end-to-end path loss, i.e., $\lvert \tilde A_n^{(k)} \rvert=1, \forall n,k$ and $\kappa^2(\Omega^{(k)})=1, \forall k$, in our proposed algorithm, respectively. 

First, it is observed from Fig.\,\ref{ChPwvsM}(a) that when $b_0=5$ or the resolution of the IRS codebook is low, all benchmarks are observed to achieve a much worse performance as compared to the proposed algorithm. The sequential update scheme even fails to output feasible paths when $M_0 < 28$. This is because its performance critically depends on the order of the update for the users. In addition, the second benchmark fails to take into account the effect of AoAs and AoDs in the network when $b_0$ is small, as discussed at the end of Section \ref{bf}; while the third benchmark overestimates the effect of CPB gain and overlooks that of end-to-end path loss. On the other hand, as $b_0$ increases to $7$, the second and third benchmarks are observed to achieve the comparable performance as the proposed algorithm when $M_0 \le 26$ and $M_0 \ge 28$, respectively. This is because the CPB gain is greatly improved with increasing $b_0$ and the effect of AoAs and AoDs diminishes. As such, the CPB gain and end-to-end path loss can dominate the BS-user effective channel gain when $M_0$ is large and small, respectively. However, when $M_0=27$, these two schemes are observed to yield worse performance than the proposed algorithm, which strikes a better trade-off between maximizing the CPB gain and minimizing the end-to-end path loss.

Finally, in Fig.\,\ref{ChPwvsBits}, we plot the max-min BS-user channel gain by the proposed algorithm and the above three benchmark schemes versus the number of controlling bits for IRS codebook in each dimension, $b_0$, under $M_0=24$. In addition, we also show the performance by the continuous IRS beamforming with $b_0 \rightarrow \infty$. It is observed that the continuous IRS beamforming yields the largest max-min BS-user channel gain among all schemes considered. This is because the maximum passive beamforming gain can be achieved at each selected IRS for any given reflection paths, i.e., $\tilde A_n^{(k)}=M, \forall n,k$. Nonetheless, as $b_0$ increases, it is observed that the performance of the proposed algorithm improves and eventually achieves a performance very close to the continuous IRS beamforming as $b_0 \ge 7$. In contrast, the sequential update scheme is observed to achieve a worse performance than our proposed algorithm as $b_0 \le 4$ and fails to output feasible paths when $b_0=6$. Although the second benchmark yields the same performance as our proposed algorithm when $b_0 \ge 6$, its performance becomes worse than ours when $b_0$ decreases. The reason is that it fails to consider the effect of AoAs and AoDs in the network as $b_0$ is small. The third benchmark is observed to achieve a worse performance than our proposed algorithm, since it overlooks the end-to-end path loss. However, it outperforms the second benchmark when $b_0 \le 3$. This indicates that the AoAs and AoDs in the network or the placement of IRSs may be more dominant than the end-to-end path-loss when $b_0$ is extremely small. All the above observations are consonant with our analysis provided at the end of Section \ref{bf}.
\begin{figure}[!t]
\centering
\includegraphics[width=3.4in]{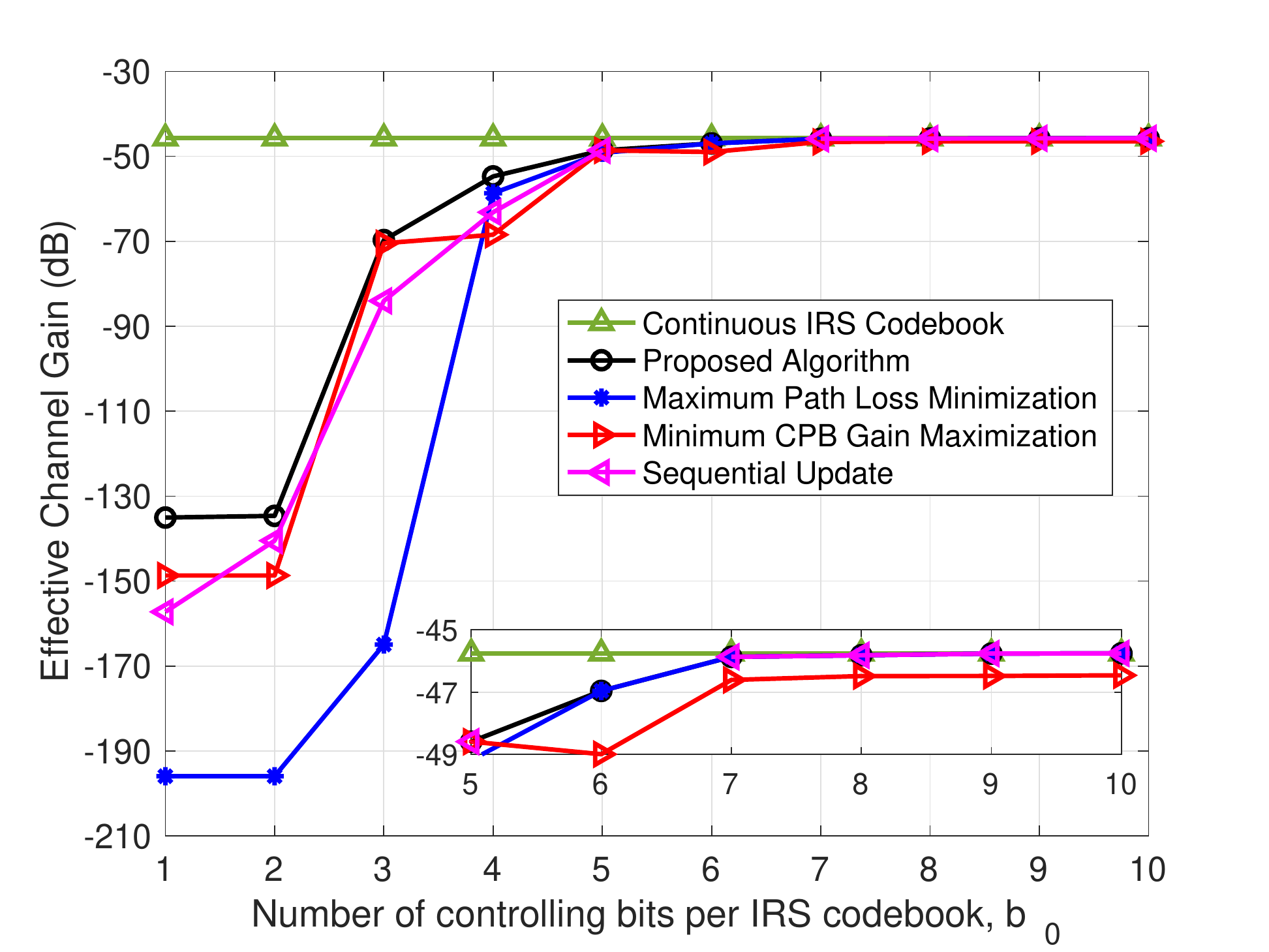}
\DeclareGraphicsExtensions.
\vspace{-6pt}
\caption{Max-min channel gain versus the number of controlling bits for IRS codebook in each dimension, $b_0$.}\label{ChPwvsBits}
\vspace{-12pt}
\end{figure}

\section{Conclusions}
This papers studies a new MBMH routing problem for a multi-IRS aided massive MIMO system, where cascaded LoS links are established between the multi-antenna BS and multiple users by exploiting the cooperative signal reflections of selected IRSs. We present the optimal active and passive beamforming solutions at the BS and each selected IRS, respectively. However, under the stringent path separation constraints for avoiding the inter-user interference, the MBMH routing problem is NP-complete and challenging to solve. To derive a high-quality suboptimal solution without incurring prohibitive complexity, we propose a parameterized recursive algorithm for this problem by leveraging graph theory. It is shown that both the number of IRS reflecting elements and size of IRS beamforming codebook can greatly impact the optimal MBMH routing solution as well as the achievable max-min BS-user channel gain. In particular, the optimal MBMH routing design should take into account the AoAs and AoDs in the system if the size/resolution of IRS beamforming codebook is not large. Besides, there exists a fundamental trade-off between minimizing the end-to-end path loss and maximizing the CPB gain, which have different dominating effects under different numbers of IRS reflecting elements.

This paper can be extended in several promising directions for future work, some of which are listed as follows to motivate future works. 
\begin{itemize}
\item First, it is interesting to study the MBMH routing problem under the general multi-path channel model. In this case, the MBMH routing problem becomes more challenging to be solved as the beamforming design cannot be simplified by assuming the LoS inter-IRS channels. In a parallel work\cite{huang2021multi}, the authors applied a deep reinforcement learning (DRL) approach to optimize the BS/IRS active/passive beamforming, under a given reflection path. As such, it is worthy of further investigating new approaches for the joint beamforming and MBMH routing design under the general multi-path channel model. Moreover, how to find an efficient approach without assuming any prior channel knowledge is challenging. 
\item  Second, the considered MBMH routing problem may become infeasible as the number of users is large or some users are close to each other in location. In this case, each IRS may be associated with more than one user to aid their transmission over orthogonal time slots or simultaneous transmission over orthogonal frequency resource blocks (RBs). In the latter case, each IRS can split its elements into multiple sub-surfaces (or co-located smaller IRSs equivalently), each associated with one user by reflecting the signal in its corresponding frequency band. As such, under any RB allocation scheme, our proposed MBMH routing design can be extended to this setup by treating the random scattering by co-located IRSs as part of environment scattering and applying our proposed path separation constraint to those IRSs reflecting user signals over the same frequency band. However, in both the cases above, user scheduling/RB allocation and IRS beam routing/passive reflection need to be jointly designed, which is interesting to study in future work.
\item Third, as our main focus is on the new MBMH routing design, we consider a simplified IRS model in this paper, while a more accurate model may be needed in practical routing design to account for other aspects pertaining to electromagnetic propagation and device hardware imperfections, such as mutual coupling among IRS elements\cite{gradoni2021end}, angle-dependent passive beamforming gain\cite{najafi2021physics}, near-field effects\cite{bjornson2020power}, etc. Nonetheless, the proposed algorithm provides a theoretical bound for the performance of practical MBMH routing designs, which can be calibrated by properly introducing correction factors to account for hardware effects by adjusting the weights in the constructed graphs accordingly.
\end{itemize}

\bibliography{IRScoop}
\bibliographystyle{IEEEtran}
\end{document}